\documentclass[12pt]{yoshida}

\usepackage{amsmath,amssymb,graphicx,color}
\usepackage{bm}
\usepackage{dcolumn}

\newcommand{\KUCPlogo}{\hbox{\lower 1.4ex\hbox{
\Huge\boldmath $\cal K$}
\kern -1.15em {\sffamily \bfseries\large\ UCP}}
\kern -4.5em \raise 0.2em\hbox{\lower 1.4ex\hbox{\color{cyan}
\Huge\boldmath $\cal K$}
\kern -1.15em {\color{magenta}\sffamily \bfseries\large\ UCP}
\put(-20,-7){\tiny\it preprint}
}}

\newcommand{\pd}{\partial}

\newcommand{\nn}{\nonumber}
\newcommand{\e}{{\rm e}}

\newcommand{\del}{\delta}
\newcommand{\ra}{\rangle}
\newcommand{\la}{\langle}
\newcommand{\rar}{\rightarrow}

\newcommand{\pdm}{\partial_{\mu}}
\newcommand{\pdn}{\partial_{\nu}}
\newcommand{\fmn}{F_{\mu\nu}}
\newcommand{\am}{A_{\mu}}
\newcommand{\an}{A_{\nu}}

\newcommand{\db}{\delta_{\rm B}}
\newcommand{\bdb}{\bar{\delta}_{\rm B}}

\newcommand{\al}{\alpha}

\newcommand{\fr}{\frac}

\renewcommand{\th}{\theta}

\newcommand{\Del}{\Delta}

\numberwithin{equation}{section}

\begin{document}

\begin{flushright}

\parbox{3.2cm}{
{KUCP-0200 \hfill \\
{\tt hep-th/0204086}}\\
\date
 }
\end{flushright}

\vspace*{0.5cm}

\begin{center}
 \Large\bf Phase Structure of Compact QED \\
from the Sine-Gordon/Massive Thirring Duality\\ 
\end{center}

\vspace*{0.7cm}

\centerline{\large Kentaroh Yoshida}

\begin{center}
\emph{Graduate School of Human and Environmental Studies,
\\ Kyoto University, Kyoto 606-8501, Japan.}
\end{center}

\centerline{\tt E-mail:~yoshida@phys.h.kyoto-u.ac.jp}

\vspace*{1.5cm}

\centerline{\bf Abstract}

We discuss a phase structure of compact QED in four dimensions 
by considering the theory as a perturbed topological model. 
In this scenario  we use the singular configuration with an appropriate 
regularization, and so obtain the results similar to the lattice gauge
theory due to the effect of topological objects. In this paper 
we calculate the thermal pressure of the topological model 
by the use of a one-dimensional Coulomb gas approximation, 
which leads to a phase structure of full compact QED.  
Furthermore the critical-line equation is explicitly evaluated. 
We also discuss relations between the monopole condensation 
in compact QED in four dimensions and the chiral symmetry 
restoration in the massive Thirring model in two dimensions.

\vspace*{1.5cm}
\noindent
Keywords:~~{\small confinement, 
monopole, vortex, Coulomb gas, sine-Gordon
model, massive Thirring model, finite temperature, phase structure}

\thispagestyle{empty}
\setcounter{page}{0}

\newpage

\tableofcontents

\newpage

\section{Introduction}

It is well known that a four-dimensional compact lattice QED 
has a confining phase at large bare coupling
 \cite{wilson,Poly,BMK,S,DQSW,OS} and hence  a linear
potential between static test charges is caused if matter
fields are discarded as dynamical variables. With decreasing bare
coupling the system changes through a phase transition to 
the Coulomb phase \cite{G} where static charges interact through the
$1/R^2$ Coulomb force at large distance $R$. We shall briefly summarize
 below. 
\begin{itemize}
 \item  \emph{ Coulomb phase}: 
The system consists of massless photons and 
diluted topological excitations which can be
interpreted as magnetic monopole currents. These monopole currents
renormalize the charge that appears in the Coulomb potential.
For small coupling monopoles are tightly bound, and the
renormalized charge nearly coincides with the bare coupling. That is,
we can almost ignore the effect of monopoles. 

\item \emph{Confining phase}:
The system consists of ``massive'' photons and
isolated monopoles, which are unbound as the coupling increases. 
The effect of monopoles play an important role in the confinement
through the dual Meissner effect and hence a linear potential is
caused \cite{NM}. Moreover, it is considered that the gauge ball (GB) state can be
formed in a confining phase, such as glue-ball in QCD. With increasing the 
coupling the renormalized charge grows until magnetic monopole 
loops unbind at the phase transition point beyond which monopoles cause the
confinement of electric charges through the dual Meissner effect.
\end{itemize}

However, a serious problem remains for more than twenty years.  
If the confining-deconfining phase transition is the second-order (the order parameter is a photon mass), then 
this limit depends on the phase we start from \cite{wilson,Peskin}. 
Thus, we cannot define the continuum limit
well on the critical-line. In order to approach this issue 
it would be useful and hopeful that we investigate  a confining phase at
strong coupling in other formulation (for an example, in our 
continuum formulation discussed in this paper).

A most well known confinement scenario is based on the monopole
condensation i.e., dual Meissner effect \cite{NM}. 
Another scenario for the quark confinement 
has been proposed by several authors \cite{kondo,HT,izawa}. 
In this scenario, full QCD$_4$ is decomposed to a perturbative
deformation (topologically trivial) part and a topological model 
(topologically non-trivial) part.
In our works \cite{KY,KY2,KY3,KY4,KY6} 
this scenario has been extended to a 
finite temperature case with zero chemical potential.
 
In the above scenario we can calculate the expectation
value of Wilson loops (at zero temperature) or
Polyakov loops (at finite temperature) from  
the topological model part. It should be remarked that the linear
potential, which means the quark confinement \cite{wilson}, can be 
explicitly derived. In this derivation the Parisi-Sourlas (PS)
 dimensional reduction \cite{PS} is powerful tool, through which  
the topological model in four dimensions is
 equivalent to a non-linear sigma model (NLSM$_2$) in two
dimensions. Thus 
calculations and considerations are essentially based on the techniques
in two dimensions. This is a great advantage of our scenario. 

This scenario is also applicable to compact QED without matter 
(a pure compact $U(1)$ gauge theory) \cite{qed}. It is well known 
in the lattice case that a confining phase exists in the 
strong-coupling regime due to monopole effects which come from 
the compactness of the $U(1)$ 
but it disappears once a continuum limit is taken (i.e., a
Coulomb phase exists.) \cite{G}.  
In this scenario singular monopole configurations (abelian
monopoles) are utilized with an appropriate regularization. 
Thus we can take
account for monopole effects and obtain the similar phase structure 
as known in the lattice gauge theory, though we consider in the {\it
continuum} formulation. That is, we can obtain in the
continuum formulation the phase structure similar to 
that in the lattice formulation. If we remove its 
regularization then the monopole effect cannot be included and 
a confining phase disappears. When we investigate the
phase structure of compact QED in our scenario we can use some 
exactly-solvable models such as two-dimensional XY model, 
sine-Gordon (SG) model and massive Thirring (MT) model. 
In particular, it has been shown in this formulation 
that a deconfining phase transition is described 
by the Berezinskii-Kosterlitz-Thouless (BKT) phase transition \cite{BKT} 
in a two-dimensional Coulomb gas (CG) system \cite{qed}. 
The behavior of the CG system decides whether the 
gauge theory is confined or deconfined. In particular, the plasma and 
molecule phases corresponds to the confined and Coulomb phases,
respectively. The above results 
have been generalized to the finite temperature case 
in Refs.\,\cite{KY,KY2,KY3}. Roughly speaking, the monopole 
dynamics of the original gauge theory in this scenario is 
effectively described in terms of vortices in an $O(2)$  
NLSM$_2$. 

In this paper we calculate the thermal pressure of the topological
model and investigate its phase structure 
which determines that of the original gauge theory. 
The thermal pressure of the
topological model can be calculated by the use of the SG/MT duality, 
and the equivalence between the one-dimensional CG system and the MT 
model with high temperature \cite{steer}. 
In conclusion we will show the phase structure of 
compact QED similar to that in the lattice gauge
theory.  
The critical-line equation is also 
explicitly evaluated, whose behavior in the high-temperature and 
strong-coupling regime coincides with that obtained in 
our previous works \cite{KY2,KY3}.  
However, the low-temperature behavior has been improved and consistent 
to the zero-temperature results \cite{qed}.

Our paper is organized as follows. In section 2, we review a deformation
of the topological model. The topological model in four
dimensions has been
mapped to an $O(2)$ NLSM$_2$ through the PS dimensional reduction. 
The relationships between parameters of the gauge theory, CG system, SG
model and MT model.
Section 3 is devoted to the equivalence 
between the one-dimensional CG and MT model with  
high temperature. A phase structure of a one-dimensional 
CG described by the thermal
pressure corresponds to  
that of compact QED in the high-temperature
and strong-coupling regime.
In section 4, we discuss a phase structure of 
compact QED from the thermal pressure of the 
topological model. Also, the
critical-line equation is explicitly evaluated. 
In section 5, the relation between the
fermion condensate in the MT model and monopole condensate in compact
QED is discussed from some exact quantities of 
a one-dimensional CG system.   
Section 6 is devoted to conclusions and discussions. 

\paragraph{\emph{Note}:}Our previous papers have contained some
misinterpretations and mistakes,
such as scale parameters, 
regularizations and numerical plots. In this
paper all of them would be sufficiently improved.

\section{Compact $\boldsymbol{U(1)}$ Gauge Theory as Deformation of Topological Model}

In this section we decompose a 
compact $U(1)$ gauge theory into a topological model (TQFT sector) 
and a deformation part. A deformation part is 
topologically trivial, but the topological model is non-trivial and 
 has topological objects such as monopoles and vortices, 
assumed to play an important role in the confinement. 
The dynamics of the confinement is encoded in the topological model, and
so we can derive a linear potential from the topological model. 
We can map the topological model to a two-dimensional $O(2)$NLSM$_2$ 
through the PS dimensional reduction \cite{PS}.
The reduced theory lives on a plane (cylinder) in the zero (finite) 
temperature case. 

The action of a (compact) $U(1)$ gauge theory on the 
$(3+1)$-dimensional Minkowski space-time is given by 
\begin{eqnarray}
 S_{\rm U(1)}[A] &=& - \fr{1}{4}\int\! d^4x\, \fmn[A] F^{\mu\nu}[A], \\
 \fmn[A] &=& \pdm\an - \pdn\am.
\end{eqnarray}
Thus, the partition function is  
\begin{eqnarray}
\label{part}
 Z_{\rm U(1)}[J]  = \int [dA][dC][d\bar{C}][dB]\,\exp\left\{i\,(S_{\rm U(1)} + S_{\rm GF+FP} + S_{\rm J})\right\},  \\
 S_{\rm J}[A,C,\bar{C},B] = \int \! d^4x \left\{J^{\mu}\am + J_{\rm c}C + J_{\rm \bar{C}}\bar{C} + J_{\rm B}B\right\}.
\end{eqnarray}
We use the BRST (Becchi-Rouet-Stora-Tyutin) quantization. Incorporating the
(anti) FP (Faddeev-Popov) ghost field $C \,(\bar{C})$ and the auxiliary field $B$, we can
construct the BRST transformation $\db$,
\begin{eqnarray}
 & & \db \am = \pdm C,~~~~\db C = 0, \nn \\
 & & \db \bar{C} = iB,~~~~\db B =0.  
\end{eqnarray}
The gauge fixing term can be constructed from the BRST transformation
$\db$ as 
\begin{eqnarray}
 S_{\rm GF+FP}[A,C,\bar{C},B]\; =\; -i\db \int \!d^4x\, G_{\rm GF + FP}[A,C,\bar{C},B],
\end{eqnarray}
and $G_{\rm GF + FP}$ is chosen as 
\begin{eqnarray}
\label{2.8}
 G_{\rm GF + FP}[A,C,\bar{C},B] \;=\; \bdb \left[\fr{1}{2}\am A^{\mu} + iC \bar{C}\right],
\end{eqnarray}
where  $\bdb$ is the anti-BRST transformation defined by 
\begin{eqnarray}
& & \bdb \am = \pdm \bar{C},~~~~\bdb C = i\bar{B}, \nn \\
& & \bdb \bar{C} = 0,~~~~\bdb \bar{B} = 0,~~~~B + \bar{B} = 0. 
\end{eqnarray}
The above gauge fixing condition (\ref{2.8}) is 
convenient to investigate the topological model. 
We decompose a gauge field as 
\begin{eqnarray}
\label{dec}
 \am  &=& V_{\mu} + \omega_{\mu}\; \equiv V_{\mu}^{U},\nn\\
\omega_{\mu} & \equiv& \fr{i}{g}\,U\pdm U^{\dagger},
\end{eqnarray} 
where $g$ is the gauge coupling constant.
Using the FP determinant $\Delta_{\rm FP}[A]$ we obtain the 
following unity 
\begin{eqnarray}
 1 &=& \Delta_{\rm FP}[A]\int [dU]\prod_{x}\del\left(\pd^{\mu}\am^{U^{-1}}\right) \nn \\
&=& \Delta_{\rm FP}[A^{U^{-1}}]\int [dU]\prod_{x}\del\left(\pd^{\mu}\am^{U^{-1}}\right) \nn \\
&=& \Del_{\rm FP}[V] \int [dU]\prod_x \del\left(\,\pd^{\mu}V_{\mu}\,\right) \nn \\
&=& \int [dU][d\gamma][d\bar{\gamma}][d\beta]\,\exp\left[\,i\int \!d^4x\left\{
\beta \pd^{\mu}V_{\mu} + i\bar{\gamma}\pd^{\mu}\pdm\gamma\, \right\}\, \right] \nn \\
 &\equiv&  \int [dU][d\gamma][d\bar{\gamma}][d\beta]\,\exp\left[\,i\int\! d^4x \left\{-i\tilde{\delta}_{\rm B}\tilde{G}_{\rm GF + FP}[V_{\mu},\gamma,\bar{\gamma},\beta]\,\right\}\,\right],
\label{unity}
\end{eqnarray}
where we have defined the new BRST transformation $\tilde{\del}_{\rm B}$ as 
\begin{eqnarray}
 & & \tilde{\del}_{\rm B}V_{\mu} = \pdm\gamma,~~~~\tilde{\del}_{\rm B}\gamma = 0, \nn \\
& & \tilde{\del}_{\rm B}\bar{\gamma} = i\beta,~~~~\tilde{\del}_{\rm B}\beta = 0.
\end{eqnarray}
In order to fix the gauge degree freedom of $V_{\mu}$, Eq.\,(\ref{unity})
is necessary. 
When Eq.\,(\ref{unity}) is inserted, the partition function can be
rewritten as follows,
\begin{eqnarray}
 Z_{\rm U(1)}[J] &=& \int [dU][dC][d\bar{C}][dB]\,\exp\{\,iS_{\rm TQFT}[\omega,C,\bar{C},B] \nn \\
& &  + \,\,iW[U;J]\,
 + \,J^{\mu}\omega_{\mu}\, + \, J_{\rm C}C \, + \,J_{\rm \bar{C}}\bar{C}\, + \,J_{\rm B}B \},  \\
S_{\rm TQFT}[U,C,\bar{C},B] &\equiv& -i\db\bdb\int\! d^4x \left\{\frac{1}{2}\omega_{\mu}^2 + iC\bar{C}\right\},
\end{eqnarray}
where
\begin{eqnarray}
\label{2.14}
 \e^{iW[U;J]} \;\equiv\; \int [dV][d\gamma][d\bar{\gamma}][d\beta]\,\exp\left\{\,
iS_{\rm pU(1)}[V,\gamma,\bar{\gamma},\beta] + i\int\! d^4x\, V_{\mu}\mathcal{J}^{\mu}\right\}, \\
\label{2.15}
S_{\rm pU(1)}[V,\gamma,\bar{\gamma},\beta] \;=\; \int\! d^4x \left\{
-\fr{1}{4}F_{\mu\nu}[V]F^{\mu\nu}[V] - i \tilde{\del}_{\rm B} \tilde{G}_{\rm GF + FP}[V,\gamma,\bar{\gamma},\beta]\right\}, \\
 \mathcal{J}_{\mu} \,\equiv\, J_{\mu} + i\db\bdb\, \omega_{\mu}. \hspace{6.0cm}\nn
\end{eqnarray}
The action (\ref{2.15}) 
describes a perturbative 
deformation part that reproduces the well-known results in ordinary
perturbation theories. 
The action $S_{\rm TQFT}$ is $\db$-exact and 
describes the topological model in which the  
information of the confinement is encoded.

In what follows we are interested in the finite-temperature
system (i.e., the system coupled to the thermal bath). 
Hence we must perform a Wick rotation of 
a time axis and move from Minkowski formulation to Euclidean one.

\paragraph{Expectation Values}

We can define the vacuum expectation value (VEV) 
in each sector using the action
$S_{\rm pU(1)}$ and $S_{\rm TQFT}$. The VEV of 
Wilson loops or Polyakov loops is an important quantity to
distinct the confinement.  In the zero-temperature case, one shall
consider the Wilson loop $W_{C}$, and its VEV satisfies 
the following relation \cite{qed}
\begin{eqnarray}
\label{expect}
 \la\, W_{C}[A]\, \ra_{\rm U(1)} &=& \la\, W_{C}[\omega]\, \la\, 
W_{C}[V] \,\ra_{\rm pU(1)} \,\ra_{\rm \scriptscriptstyle{TQFT}} \nn\\
& =& \la\, W_{C}[\omega]\, \ra_{\rm \scriptscriptstyle{TQFT}}\,\la\, W_{C}[V]\, \ra_{\rm pU(1)},
\end{eqnarray}
where the contour $C$ is rectangular. 
The VEV of a Wilson loop is completely
separated into the TQFT sector and a perturbative deformation part.
That is, we can evaluate the VEV in the TQFT sector 
independently of the perturbative deformation part. In fact, 
we can derive the linear potential by investigating the TQFT
sector.  

\begin{figure}
\centerline{\includegraphics{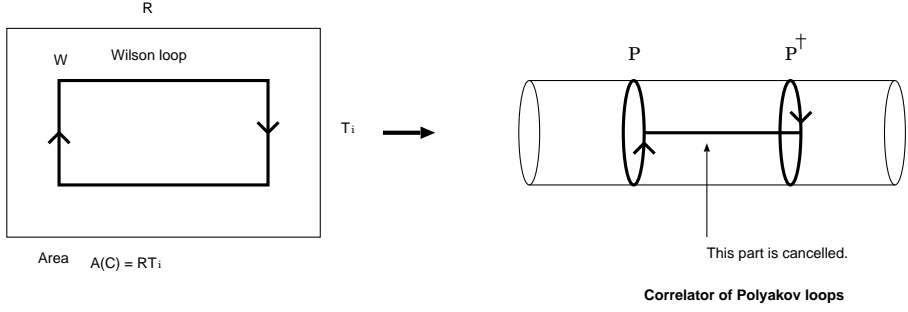}}
\caption{\footnotesize\label{poly:fig}
The correlator of Polyakov loops. The expectation
value of a Wilson loop $W$ is equivalent to a correlator of the
Polyakov loops $P$ and $P^{\dagger}$.}
\end{figure}
In the finite-temperature case, 
we must evaluate the correlator of Polyakov
loops $P({\bf x})$. It can be evaluated in the same way as the Wilson loop, 
due to the following relation (as shown in Fig.\ref{poly:fig})
\begin{equation}
 \la \,P({\bf x}) P^{\dagger}(0)\, \ra_{\rm U(1)} = \la\, W_{C}[A]\,\ra_{\rm U(1)}. 
\end{equation}

Furthermore we can derive a Coulomb potential (at zero temperature) \cite{qed}
or Yukawa-type potential (at finite temperature) using a hard 
thermal-loop (HTL) approximation \cite{KY2} from a perturbative deformation part.

\section{Topological Model and Two-dimensional Systems}

\subsection{Topological Model and Parisi-Sourlas Dimensional Reduction}

The action of the topological model in four dimensions, 
\begin{equation}
 S_{\rm TQFT}[U,C,\bar{C},B] \;=\; \db\bdb\int\! d^4 x \,\left\{\frac{1}{2}\omega_{\mu}^2 + iC\bar{C}\right\}, 
\end{equation}
can be rewritten through the PS dimensional reduction \cite{PS} 
as 
\begin{eqnarray}
 S_{\rm TQFT}[U] &=& \pi \int \! d^2x\, \omega_{\mu}^2,\qquad\omega_{\mu} \equiv \frac{i}{g}U(x)\pdm U(x)^{\dagger}   \nn \\
&=& \frac{\pi}{g^2}\int\! d^2x \,\pdm U\, \pdm U^{\dagger}, 
\label{ac-nlsm}
\end{eqnarray}
where we have omitted the ghost term. The above action (\ref{ac-nlsm}) describes an
$O(2)$ NLSM$_2$.  
Using $U(x) = \exp \{i\varphi(x)\}$ we obtain 
\begin{equation}
\label{2.22}
 S_{\rm TQFT}[\varphi] \;=\; \frac{\pi}{g^2}\int\! d^2x\, (\pdm \varphi)^2.
\end{equation}
If the $U(1)$ is not compact, then it  
becomes a free scalar field theory in two dimensions and 
has no topological object. Thus a confining phase cannot exist. 
If the $U(1)$ is compact, then it describes 
a periodic boson theory. The angle variable
$\varphi(x)$ is periodic (mod $2\pi$), and so  $\varphi(x)$ 
is a compact variable. It is well known that 
a compactness plays an important role in the confinement \cite{Poly}.
When we consider the system at finite (zero) temperature, 
the dimensionally reduced theory lives on the cylinder (plane). 

Let us consider the zero-temperature case. The solution of the 
classical field equation $\nabla^2\varphi(x) = 0\;({\rm mod}~2\pi)$ is a
harmonic function and hence $\varphi(x)$ is either constant or has
singularities if we require that $\varphi(x)$ is constant at infinity. 
The solutions are 
called vortices and expressed by 
\begin{equation}
 \varphi(z) = \sum_{i=1}^{n}Q_{i}\,{\rm Im}\,\ln(z - z_{i}),~~~z\equiv x_1 + i\,x_0,
\end{equation}
where $Q_i$ are the intensities of vortices and 
isolated singularities only are assumed.  
We can easily see that $\varphi(x)$ is a multi-valued function and 
varies by $2\pi$ as one turns a singularity anti-clockwise, but
$\exp\{i\varphi(x)\}$ is well-defined everywhere, 
except at the singular points.
The classical action for such a singular configuration is infinite, 
one can regularize the action by excluding a small region
around each singularity (for example, a disk of radius $R_0$ centered on
each singular point).  Let $D$ be the remaining integration domain, we
can obtain 
\begin{eqnarray}
\label{CG-vol}
S_{\rm vor}\left[\,(Q_j,z_j)\,\right] &\equiv& \frac{\beta'}{2}\int_{D}d^2x \,
(\pdm\varphi)^2, \quad \beta' \equiv \frac{2\pi}{g^2} \nn \\
&=& - \frac{8\pi^3}{g^2}\sum_{i < j}\,\frac{Q_i Q_j}{2\pi}\,\ln |z_i - z_j|\, +\, \sum_{i=1}^{n}Q_i^2 \,\frac{2\pi^2}{g^2}\,\ln\frac{1}{R_0}.
\end{eqnarray}

\begin{figure}
 \begin{center}
  \includegraphics[scale=.6]{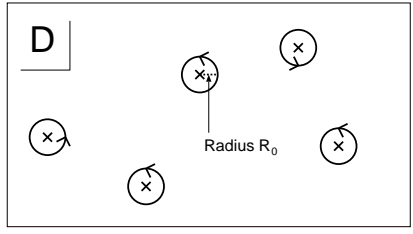}
 \end{center}
\caption{\footnotesize Vortices and integration domain $D$.}
\label{vortex:fig}
\end{figure}
When we decompose the variable $\varphi$ as 
\begin{equation}
 \varphi \;=\; \varphi_{\rm vor}\, +\, \varphi_{\rm fluc}
\end{equation}
where $\varphi_{\rm vor}$ is the vortex configuration and $\varphi_{\rm
fluc}$ denotes the fluctuation around the classical configuration i.e., 
the spin wave,  
the action of an $O(2)$ NLSM$_2$ can be rewritten near a stationary
configuration as 
\begin{eqnarray}
 S_{\rm TQFT} = S[\,(Q_i,z_i)\,] + \frac{\beta'}{2}\int\!d^2x(\pdm\varphi_{\rm fluc})^2,
\end{eqnarray}
and summing over all vortex sectors yields an approximate partition function 
\begin{eqnarray}
 Z_{\rm TQFT} = Z_{\rm SW}\cdot Z_{\rm CG}.
\end{eqnarray} 
The partition function of a CG system $Z_{\rm CG}$ is given by 
\begin{eqnarray}
 Z_{\rm CG} &=& \sum_{n=0}^{\infty}\frac{\zeta^{2n}}{(n!)^2}\prod_{i=1}^{n}\int \frac{d^2x_i}{a^2}\frac{d^2y_i}{a^2}\, \exp\Bigg[- \frac{8\pi^3}{g^2}\,\Big\{
\sum_{i < j}\left[\>\Delta(x_i - x_j) + \Delta(y_i - y_j)\right] \nn \\
& & - \sum_{i,j}\,\Delta(x_i - y_j)  \, \Big\} \,\Bigg],
\label{CG-gauge}
\end{eqnarray}
where the temperature of the CG system is  
\begin{equation}
\label{3.8}
 T_{\rm CG} \equiv \frac{g^2}{8\pi^3},
\end{equation}
and $\Delta(x_i - x_j)$ expresses the Coulomb potential on the plane, 
\begin{equation}
\label{prop-zero}
 \Delta(x - y) = -\frac{1}{2\pi}\ln\frac{|x - y|}{a}.
\end{equation}
Here we have introduced the cut-off $a$ for the short-range interaction
(length scale)
through the change of variable $x_i \rar x_i/a$ and 
\begin{equation}
\zeta \equiv \exp\left[\frac{2\pi^2}{g^2}\,\ln \frac{R_{0}}{a}\right] 
\equiv \,\kappa^{2\pi^2/g^2},\quad \kappa \equiv \frac{R_0}{a},
\end{equation}
plays a role of the chemical potential of the CG system and 
is determined by the value
of $R_0$. The partition function of the spin wave $Z_{\rm SW}$ 
include no topological
information and hence cannot contribute to the confinement in the gauge
theory. Therefore we ignore this part hereafter. 

\begin{figure}
\centerline{\includegraphics[scale=.7]{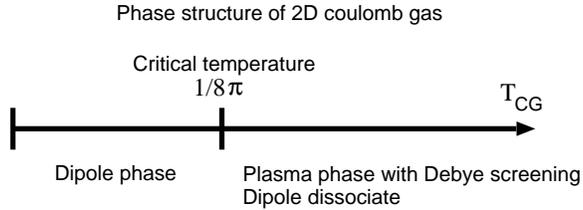}}
\caption{\footnotesize Two-dimensional Coulomb gas has two phases. Over the
critical temperature $T_{\rm CG} =1/8\pi$, the system is in a plasma phase with
Debye screening, and so the mass gap exists. Below $T_{\rm CG}$ it is in
 a dipole phase, in which Coulomb charges form dipoles. The system
has a long range correlation and no mass gap.}
\label{CG:fig}
\end{figure}

The CG system undergoes the BKT phase transition at certain temperature
$1/8\pi$  
as shown in Fig.\ref{CG:fig}. 
In this case, the critical temperature $T_{\rm CG} = 1/8\pi$ implies the
critical coupling $g_{\rm cr} = \pi$ (see Appendix \ref{BKT:app} for
details). If $g < \pi$, then vortex and anti-vortex are tightly bound and 
a neutral pair of vortices tends to be formed. Thus the vortex effect
can be ignored and hence the gauge theory is in the Coulomb
phase. While, if $g > \pi $, then the CG system is in the plasma phase
and hence the vortex effect exists. 
Therefore the linear potential between test charged
particles is induced by vortices and the system is confined.  
The string tension can be obtained from 
 the expectation value of the Wilson loop 
(for details, see Ref.\,\cite{qed}),
\begin{eqnarray}
 \la\,W_C[A]\,\ra_{\rm U(1)} &=& \la\,W_C[V]\,\ra_{\rm pU(1)}\,\la\,W_C[\omega]\,\ra_{\rm \scriptscriptstyle{TQFT}}, \nn \\
\la\,W_C[\omega]\,\ra_{\rm \scriptscriptstyle{TQFT}} &\cong& \e^{-\sigma_{\rm st}A(C)}, \\
\sigma_{\rm st} &=& 4\pi^2\left(\frac{e}{g}\right)^2\zeta\Lambda^2 \;=\; 4\pi^2\left(\frac{e}{g}\right)^2\kappa^{2\pi^2/g^2}\, \Lambda^2 \nn \\
&=&  4\pi^2\left(\frac{e}{g}\right)^2\left[\frac{R_0}{a}\right]^{2\pi^2/g^2}\Lambda^2
\end{eqnarray}
where $A(C) = R\,T_{i},\, R=|{\bf x}|$ and $e$ is a charge of test
particles. In particular, if the particles belong to 
the fundamental representation, $e = g/2$. 
Whether a confining string vanishes or not is determined by 
the phase of the CG system. Note that a string
tension $\sigma_{\rm st}$ is proportional to $\zeta$. The $\zeta \rar 0$ 
limit means $R_0 \rar 0$ and implies that the regularization of the
action for the vortex configuration. Therefore the action tends to
infinity 
and we fail to include the effect of vortices. As a result, 
a confining phase also disappears in the limit $\zeta\rar 0$ as expected.

\paragraph{Comments on the relations between monopole in four dimensions
    and vortices in two-dimensions}
From the viewpoint of the four-dimensional gauge theory, 
vortices in the two-dimensional $O(2)$ NLSM$_2$ would be interpreted 
as the point that the monopole world-line pieces the plane chosen 
when we use the PS mechanism as discussed in
Refs. \cite{kondo,qed}. (Concretely discussed in the $SU(2)$ case. The analogy
holds in the $U(1)$ case.) In particular, a singularity (regularization) 
of an abelian monopole in the $U(1)$ theory 
should imply that of a vortex in the $O(2)$NLSM$_2$. It would be also
expected that parameters $R_0$
and $a$ are a radius and a short-range cut-off for monopoles,
respectively.  

The effect of monopoles which is relevant to the confinement 
of the system should be limited on the plane (area) enclosed 
by the Wilson loop, and we would be able to study 
the monopole effect effectively 
as vortex effect in the two-dimensional theory in our scenario through
the PS mechanism.   
Technically, the gauge transformation $U(x)$ has the topological nature 
of the system or gauge group (i.e., monopole configuration), 
and we are considering the dynamics of this part 
in the two-dimensional theory through the PS dimensional reduction.  
Thus, we could consider the nature of the phase transition 
in the four-dimensional gauge theory as that of the CG system 
in the two-dimensional theory. 
One should consider that the phase transition from the viewpoint of 
four dimensions has a BKT-nature, which would be identical 
with that of the one-dimensional CG system.

\subsection{Equivalence between Sine-Gordon model and Coulomb Gas System}

The Euclidean action of the  SG model on the plane (at zero temperature) or 
the cylinder (at finite temperature) is defined by  
\begin{equation}
S_{\rm SG}[\phi] \;=\;
\int\! d^2x \left\{\frac{1}{2}\left(\pdm \phi\right)^2 - \frac{m^2}{\bar{\lambda}}
\left[\cos\left(\sqrt{\bar{\lambda}}\phi\right) - 1\right]\right\}.
\end{equation}
The above action contains two parameters $\bar{\lambda} \equiv \lambda
/m^2$ and $m$. The 
$\bar{\lambda}$ is a dimensionless parameter but $m$ is dimensionful one. 
The partition function is given by 
\begin{eqnarray}
 Z_{\rm SG} &=& \int [d\phi] \,\exp \left\{- S_{\rm SG}\right\}.
\end{eqnarray}
This can be rewritten as follows,
\begin{eqnarray}
 Z_{\rm SG} &=& 
\sum_{n=0}^{\infty}\frac{1}{n!}\left(\frac{m^2}{\bar{\lambda}}\right)^n
 \int [d\phi]\, \e^{-\int d^2x (1/2)\left(\pdm \phi\right)^2  } \left[\int\! d^2x \,\cos\left(\sqrt{\bar{\lambda}}\phi\right)\right]^n \nn \\
&=& \sum_{n=0}^{\infty}\frac{1}{(2n)!}\left(\frac{m^2}{2\bar{\lambda}}\right)^{2n} \int [d\phi]\, \e^{-\int d^2x (1/2)\left(\pdm \phi\right)^2  } \left[\int\! d^2x\left\{ \e^{i\sqrt{\bar{\lambda}}\phi} + \e^{-i\sqrt{\bar{\lambda}}\phi}\right\} \right]^{2n} \nn \\
&=& \sum_{n=0}^{\infty}\frac{1}{(n!)^2}\left(\frac{m^2}{2\bar{\lambda}}\right)^{2n}   \int [d\phi]\, \e^{ - \int d^2x
 (1/2)\left(\pdm \phi\right)^2}\prod_{j=1}^{2n}\int\! d^2x_j\,\e^{iq_j\phi(x_j)} 
\nn \\
&=& \sum_{n=0}^{\infty}\frac{1}{(n!)^2}\left(\frac{m^2}{2\bar{\lambda}}\right)^{2n}  \left\la\prod_{j=1}^{2n}\int d^2x_j\,\e^{iq_j\phi(x_j)}\right\ra,
\end{eqnarray}
where we have defined 
\begin{equation}
q_i = \sqrt{\bar{\lambda}}\,\epsilon_i  = 
\left\{
\begin{array}{cc}
\sqrt{\bar{\lambda}} & {\rm for}~ 1 \leq i \leq n,  \\
-\sqrt{\bar{\lambda}} & ~~~~~~~~{\rm for}~n + 1 \leq i \leq 2n. 
\end{array}
\right.
\end{equation}
We have omitted the term $\exp\left[- (m^2/\bar{\lambda})\int d^2x\right]$ since it
is canceled by the normalization of the partition function. 
Here we note that 
\begin{eqnarray}
\left\la \e^{\int d^2x\, J\phi} \right\ra &\equiv& \int [d\phi]\exp
\left[\int \!d^2x \left(-\frac{1}{2}\left(\pdm\phi\right)^2 + J\phi\right)\right] \nn \\
&=& \exp \left[\iint\! d^2x d^2y\, J(x)\Delta (x-y) J(y)\right],
\label{1.3}
\end{eqnarray}
where $\Delta(x-y)$ is a massless scalar field propagator,
\begin{equation}
 \Delta(x-y) \;\equiv\; \int\! \frac{d^2p}{(2\pi)^2}\,\frac{\e^{i p\cdot(x-y)}}{p^2},
\label{1.4}
\end{equation}
and this expression leads to Eq.\,(\ref{prop-zero}) 
if we consider at zero temperature. 
In particular, if 
we choose an external field $J(x)$ as 
\begin{equation}
J(x) \;=\; i\sum_{i=1}^{n} \,q_i\, \del(x-x_i),
\end{equation}
then we obtain  
\begin{equation}
\left\la \prod_{j=1}^{2n} \e^{iq_j \phi(x_j)}  \right\ra \;=\; 
\left\{
\begin{array}{cc}
\exp\left[- (\bar{\lambda}/2)\sum_{j,k}^{~~}\epsilon_j \epsilon_k\Delta(x_j - x_k)\right] & {\rm for}~\sum_{i=1}^{2n}\epsilon_i = 0,  \\
0 & {\rm for}~\sum_{i=1}^{2n}\epsilon_i \neq 0. 
\label{for}
\end{array}
\right.
\end{equation}
If the net charge is non-zero, then correlation functions vanish 
due to the translation symmetry under the following transformation 
\[
\phi(x) \;\longrightarrow\; \phi(x) \;+\;{\rm const}.
\]
Using Eq.\,(\ref{1.3}), we obtain the following partition function, 
\begin{eqnarray}
Z_{\rm SG}\! &=& \!\sum_{n=0}^{\infty}\frac{1}{(n!)^2}\left(\frac{m^2}{2\bar{\lambda}}\right)^{2n}\prod_{i=1}^{n}\int\! d^2 x_i d^2y_i\, \exp \Bigg[- 
\bar{\lambda} 
\Big\{ \sum_{i < j }[\Delta (x_i - x_j) + \Delta (y_i - y_j)] \nn \\
& & - \sum_{i,j} \Delta(x_i - y_j) \,\Big\}\,\Bigg], 
\label{CG-SG}
\end{eqnarray}
where coordinates $x_i,~(1\leq i \leq n)$ indicate positions of vortices 
and new coordinates $y_i \equiv x_{n+i},~(1\leq i \leq n)$ denote those of
anti-vortices. 
Thus it has been shown that the SG model is equivalent 
to a neutral CG system whose temperature is defined by
\begin{equation}
 T_{\rm CG} \,=\, \frac{1}{\bar{\lambda}}. 
\end{equation}
The phase transition in the SG model at the critical
coupling $\bar{\lambda} = 8\pi$ (often called the Coleman
transition)  
corresponds to the BKT phase transition at the temperature $T_{\rm CG} =
1/8\pi$ in terms of the CG system. 

One can identify two CG systems (\ref{CG-gauge}) and
(\ref{CG-SG}) through the following relations,   
\begin{eqnarray}
\frac{g^2}{8\pi^3} \, = \,\frac{1}{\bar{\lambda}}\quad (\mbox{temperature}), \qquad
\frac{\zeta}{a^2} \,=\, \frac{m^2}{2\bar{\lambda}}\quad (\mbox{chemical potential}).
\end{eqnarray}
Thus, parameters of the SG model can be expressed 
by those of compact QED as 
\begin{equation}
\label{relation}
 \bar{\lambda} \,=\, \frac{8\pi^3}{g^2},\qquad m \,=\, 
\frac{4\pi^{3/2}}{g}\,\zeta^{1/2}\Lambda,\qquad \Lambda \equiv \frac{1}{a}.
\end{equation} 
It should be noted that the length scale 
of the CG system, $a$ gives mass scale in the SG model.  
Recall that a string tension is $\sigma \,\propto\, \zeta$, 
and so vanishes as
$\zeta \rar 0~ (R_0 \rar 0)$ and a confining phase disappears. 
The limit $\zeta\rar 0$, which removes the regularization 
for singular vortices, 
means $m^4/\lambda \rar 0$ 
and the SG model becomes free scalar field theory. That is, 
no kink solution exists.  Kinks in the SG model 
are closely related to the existence of vortices in the CG
system. Physically, this limit means that 
the fugacity of vortices (monopoles) becomes zero.  
Consequently, the effects of topological objects vanishes 
and hence a confining phase disappears. 

It is also commented that the variable $\varphi$ in an $O(2)$ NLSM$_2$ 
is related to $\phi$ in the SG model through 
\begin{equation}
\label{o2-sg}
 \pdm \varphi \,=\, \frac{(2\pi)^{3/2}}{g}\,\epsilon_{\mu\nu}\pdn\phi.
\end{equation}

\begin{figure}
\centerline{\includegraphics[scale=.8]{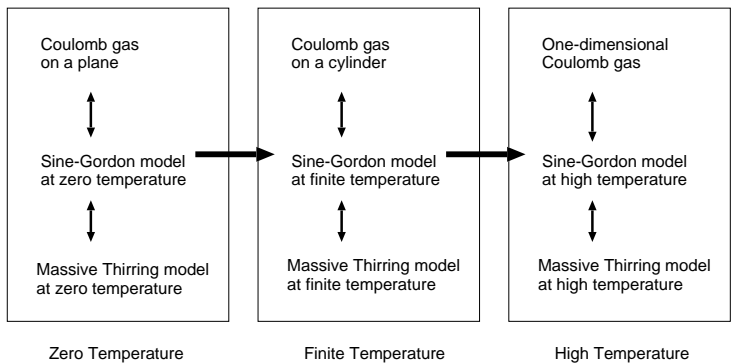}}
\caption{\footnotesize\label{tqft:fig}Relations between the CG system,
 SG model and MT model at zero and finite temperature.}
\end{figure}

It is well known that the SG model is equivalent to the MT model. 
We shall summarize the relations between the CG system, SG and MT
model in Fig.\,\ref{tqft:fig}. It should be noted that 
the equivalence holds even in the finite-temperature case 
\cite{t-boson,steer2}.
We shall also list phase structures of these models on a 
plane 
in Tab.\,\ref{phase:tab}.  
\begin{table}
\begin{center}
{\footnotesize 
\begin{tabular}{|c|c|c|c|}
\hline
 & Coulomb Gas  & Sine-Gordon & Massive Thirring \\
\hline\hline
 & $T_{\rm CG} > 1/8\pi$ & $0 < \bar{\lambda} < 8\pi $    & $g_{\rm \scriptscriptstyle{MT}}^2 > - \pi/2$   \\
\hline
Correlation Length & finite (Debye screening) & finite (mass scale) &
  finite (mass scale) \\
\hline
Phase & plasma & $\cos$ potential:  & massive \\
 & (vortices effect exists) & relevant  & relevant  \\ 
\hline
Renormalizability & & super-renormalizable & super-renormalizable \\
\hline
Peculiar Point 
& & $\bar{\lambda}=4\pi$ & $g_{\rm \scriptscriptstyle{MT}}^2 =0$   \\
 & & free massive scalar$^{\dagger}$ & free massive fermion  \\
\hline\hline
Transition Point & $T_{\rm CG}=1/8\pi$ & $\bar{\lambda}= 8\pi$ & $g_{\rm \scriptscriptstyle{MT}}^2= -\pi/2$ \\
\hline
  &   &  $\cos$ potential: &  \\
 &  &  marginally irrelevant & marginally irrelevant \\
\hline
Renormalizability &  & renormalizable & renormalizable \\
\hline\hline 
  & $T_{\rm CG} < 1/8\pi$ & $\bar{\lambda} > 8\pi$ & $g_{\rm \scriptscriptstyle{MT}}^2 < -\pi/2$ \\
\hline
Correlation Length & infinite (no screening) & infinite &  infinite \\
  &  & no quantum ground state & \\
\hline
Phase & molecule  & $\cos$ potential: &  massless \\
 & (no vortex effect) & irrelevant & irrelevant  \\
\hline
Renormalizability & & not renormalizable & not renormalizable \\   
\hline
 \end{tabular}
}
\end{center}
\vspace{0.2cm}

{\scriptsize $^\dagger$: Additional renormalization is needed.}
\caption{\footnotesize Phases of the Coulomb gas system,
 sine-Gordon  and massive Thirring model.}
\label{phase:tab}
\end{table}

\paragraph{Confining Phase Transition from the viewpoint of the 
Renormalization Group}

From the perturbed conformal field theory (CFT) point of view, the SG
model can be considered as a perturbation of the free massless
two-dimensional scalar field theory by the relevant conformal operators
$\exp(\pm i \sqrt{\bar{\lambda}}\phi)$ whose scale dimensions
\begin{equation}
 \Delta_{\rm SG} = \frac{\bar{\lambda}}{8\pi} = \frac{p}{p + 1},
\end{equation}
where we have introduced a real parameter $p$ defined by 
\begin{equation}
 p \equiv \frac{\bar{\lambda}}{8\pi - \bar{\lambda}}.
\end{equation}
In fact, if we take $0 \leq \sqrt{\bar{\lambda}} \leq \sqrt{8\pi}$, then
there is a family of CFTs depending on one parameter $p$. 
The corresponding mass parameter $m^2$ has a complementary dimension,
\begin{equation}
 m^2 \,\sim\, [{\rm mass}]^{2/(p+1)} = [{\rm mass}]^{(8\pi - \bar{\lambda})/4\pi}.
\end{equation} 
Thus we can read off the behavior of a cosine potential for
each value of $\bar{\lambda}$. It is relevant for $0\leq \bar{\lambda}
 < 8\pi$, marginal for $\bar{\lambda}=8\pi$, and irrelevant for 
$\bar{\lambda} > 8\pi$.

\subsection{Gauge Theory and Massive Thirring Model}

The action of the MT model 
in the two-dimensional Euclidean space with metric $(+,+)$, 
is given by 
\begin{equation}
 S_{\rm MT}[\psi,\bar{\psi}] \;=\; \int\! d^2x\,\left\{-\bar{\psi}(\gamma^{\mu}\pdm + m_{\rm \scriptscriptstyle{MT}})\psi + \frac{1}{2}g_{\rm \scriptscriptstyle{MT}}^2 \bar{\psi}\gamma_{\mu}\psi\bar{\psi}\gamma^{\mu}\psi\right\}.
\end{equation}
The Euclidean gamma matrices (following the notation in Ref.\,\cite{Jinn}) are defined by 
\begin{eqnarray}
 \gamma^1 \,=\, \sigma_1 \,=\, \begin{pmatrix}
0 & 1 \\ 
1 & 0
\end{pmatrix}
, \quad \gamma^2 \,=\, \sigma_2 \,=\, \begin{pmatrix}
0 & -i \\
i & 0 
\end{pmatrix}
, \quad \gamma^5 \,=\, -i \sigma_1 \sigma_2 \,=\, \sigma_3 \,=\, 
\begin{pmatrix}
1 & 0 \\
0 & -1
\end{pmatrix}
,
\nn
\end{eqnarray}
and obeys the standard relations
\begin{eqnarray}
\gamma^{\mu}\gamma^{\nu} \,+\, \gamma^{\nu}\gamma^{\mu}\,=\, 
2\del^{\mu\nu}, & & 
\gamma^{\mu}\gamma^5 \,+\, \gamma^5\gamma^{\mu} \,=\, 0, \nn \\
\gamma^\mu\gamma^\nu \,=\, \del^{\mu\nu} \,+\, i\epsilon^{\mu\nu}\gamma^5, 
& & \gamma^{\mu}\gamma^{5} \,=\,-i\epsilon^{\mu\nu}\gamma^{\nu}, \nn
\end{eqnarray}
are satisfied. The anti-symmetric tensor $\epsilon_{\mu\nu}$ is
defined by $\epsilon_{12} = - \epsilon_{21} = 1,\; \epsilon^{12} = -
\epsilon^{21} = 1$.  

It has been well known that the SG/MT duality holds 
at zero temperature \cite{cole} and
finite temperature \cite{t-boson,steer2} if the following relations 
\begin{eqnarray}
\label{sg/mt}
\frac{4\pi}{\bar{\lambda}} \,=\, 1 + \frac{g_{\rm \scriptscriptstyle{MT}}^2}{\pi},\qquad \frac{m^2}{\bar{\lambda}} \,=\, \rho\, m_{\rm \scriptscriptstyle{MT}},
\end{eqnarray} 
are satisfied. Here $\rho$ and $m_{\rm \scriptscriptstyle{MT}}$ are a 
renormalization scale and a 
renormalized mass at that scale, respectively. Hereafter, we shall 
set $\rho = m_{\rm
\scriptscriptstyle{MT}}$. 
The condition (\ref{sg/mt}) can be rewritten as  
\begin{eqnarray}
\label{mt-sg}
m_{\rm \scriptscriptstyle{MT}} \,=\, \frac{m}{\sqrt{\bar{\lambda}}},\qquad g_{\rm \scriptscriptstyle{MT}}^2 \,=\,\frac{4\pi^2}{\bar{\lambda}} - \pi.
\end{eqnarray}
If a coupling constant of the SG model $\bar{\lambda}\equiv
\lambda/m^2$ 
becomes $4\pi$, then a mass of the MT model becomes zero. 
Thus $\bar{\lambda} = 4\pi$ corresponds to a massless
Thirring model. Combining Eq.\,
(\ref{mt-sg}) with  Eq.\,(\ref{relation}), we can obtain the relations 
between each parameter of the MT model and gauge theory as  
\begin{eqnarray}
\label{MT-gauge}
m_{\rm \scriptscriptstyle{MT}} \,=\, \sqrt{2\zeta}\Lambda,\qquad g_{\rm \scriptscriptstyle{MT}}^2 \,=\, \frac{g^2}{2\pi} - \pi.
\end{eqnarray}
Here, note that $m_{\rm \scriptscriptstyle{MT}}$ vanishes 
in the limit of $\zeta \rar 0~ (R_0 \rar 0)$. 
That is, the MT model becomes a massless Thirring model. This fact is
consistent to that a cosine potential of the SG model also vanishes as
$\zeta \rar 0$ since a mass perturbation in the MT model corresponds
to a cosine potential perturbation in the SG model. 

Also, the current and mass equivalence, respectively,  
\begin{eqnarray}
\label{current}
 \bar{\psi}\gamma_{\mu}\psi &=& \frac{\sqrt{\bar{\lambda}}}{2\pi}\,\epsilon_{\mu\nu}\pdn\phi, \\
  m_{\rm \scriptscriptstyle{MT}}\,\bar{\psi}\psi &=& \frac{m^2}{\bar{\lambda}}\cos\left(\sqrt{\bar{\lambda}}\phi\right),
\label{mass}
\end{eqnarray}
are also useful in discussing the relations between topological charges, 
such as a monopole charge in the gauge theory, vortex intensity in the
$O(2)$ NLSM$_2$, kink
number in the SG model and fermion number in the MT model.  
In particular, Eq.\,(\ref{current}) means the equivalence between a 
kink number and fermion number.
 
We can again understand a mass term in the MT model from the viewpoint
of the perturbed CFT. Recall that a cosine potential in the SG model 
is a perturbation from a massless free scalar theory. A mass
term in the action of the MT model 
corresponds to a cosine potential through the
abelian bosonization, and we can identify a mass term in the MT
model as a perturbation from a massless Thirring model. Correspondingly, 
the monopole effect realized by a cosine potential in the SG model is  
described by a mass term in the MT model. Such a correspondence leads
us to expect the relation between the monopole condensate 
in compact QED and fermion one in the MT model.

\paragraph{Relations of Some Quantities}

\begin{table}[htb]
\begin{center}
{\footnotesize
 \begin{tabular}{|c|c|c|c|c|}
\hline
  & Gauge  &  $O(2)$ NLSM$_2$     &      SG model     &  MT model     \\
 \hline\hline 
Variable & $\am = V_{\mu} + \omega_{\mu}$ & $U = \e^{i\varphi}$ & $\phi$ & $\psi$ \\
\hline 
Topological Objects & monopoles & vortices & kinks  & fermions \\
\hline
Topological Charge & magnetic charge   & intensity  & kink number &
  fermion number  \\
 & & & & $U(1)_V$ \\
\hline

Global Symmetry & $\e^{i\al}\in U(1)$ &
  $\varphi\rar\varphi + 2\pi n$ & $\phi\rar\phi + \frac{2\pi}{\sqrt{\bar{\lambda}}}n$ & $\psi\rar \e^{2\pi n \gamma^5}\psi$ \\
(Residual)  & global $U(1)$ & $^{\forall}n\in \mathbb{Z}$ & $^{\forall}n\in\mathbb{Z}$ & $^{\forall}n\in \mathbb{Z}$ \\
\hline
Broken Symmetry & & $\varphi\rar\varphi + \alpha$ & $\phi\rar \phi + \frac{1}{\sqrt{\bar{\lambda}}}a$ & $\psi\rar\e^{ia\gamma^5}\psi$ \\
by Topological Objects & & $^{\forall}\e^{i\alpha}\in U(1)$ & $ ^{\forall}a\in \mathbb{R}$  & $^{\forall}\e^{ia\gamma^5}\in U(1)_{A}$ \\
\hline
Scale Parameters
&   \multicolumn{1}{@{}c@{}|}{%
\begin{tabular}{c}
$\Lambda \equiv 1/a$ \\ 
UV cut-off$^{\dagger 1}$ \\
\hline
$R_0$ \\
monopole radius
\end{tabular}
}
&   \multicolumn{1}{@{}c@{}|}{%
\begin{tabular}{c}
$a$ \\ 
length scale$^{\dagger 2}$ \\
\hline
$R_0$ \\
vortex radius
\end{tabular}
}

& 
\multicolumn{1}{@{}c@{}|}{%
\begin{tabular}{c}
$m = \frac{4\pi^{3/2}}{g}\zeta^{1/2}\Lambda$ \\ 
mass 
\end{tabular}
}

& \multicolumn{1}{@{}c@{}|}{%
\begin{tabular}{c}
$m_{\rm \scriptscriptstyle{MT}} = \sqrt{2\zeta}\Lambda$ \\ 
fermion mass
\end{tabular}
}
 
\\
\hline
\multicolumn{1}{|@{}c@{}|}{%
\begin{tabular}{c}
Characteristic \\
Parameter
\end{tabular}
}  & $g,\,a,\,R_0$  & $\zeta,\,a,\, R_0$  &
  $m,\,\bar{\lambda}$   &  $m_{\rm \scriptscriptstyle{MT}},\,g_{\rm \scriptscriptstyle{MT}}$
  \\
\hline
Source  & $\pdm\omega_{\nu}$  & $\frac{1}{g}\pdm \varphi$  &
  $ \frac{\sqrt{\bar{\lambda}}}{2\pi}\epsilon_{\mu\nu}\pdn\phi$  & $j_{\mu}\equiv \bar{\psi}\gamma_{\mu}
\psi$  \\
 & magnetic source &  & kink number  & vector current \\
\hline
(Dual) & $\epsilon_{\mu\nu\rho\sigma}\pd_{\rho}\omega_{\sigma}$  &
  $\frac{1}{g}\epsilon_{\mu\nu}\pdn\varphi$  &
  $\frac{\sqrt{\bar{\lambda}}}{2\pi}\pdm\phi$  & $j_{5\mu}\equiv \bar{\psi}\gamma_{\mu}\gamma_5\psi$ \\
Source & electric source &     &   & axial current \\
\hline
If non-compact & no monopole  & no vortex & free boson & massless fermion \\
\hline
\end{tabular}
} 
\end{center}

{\scriptsize $\dagger 1$, $\dagger 2$\,: The parameter $a$ denotes short-range cut-off for interaction between
 monopoles or vortices, respectively.}

\caption{\footnotesize Relations of some quantities. }
\label{list:tab}
\end{table}

We shall briefly remark that some quantities such as
currents, topological objects and these charges are closely related.     
Using Eq.\,(\ref{o2-sg}) and Eq.\,(\ref{current}) leads to 
the current relation,   
\begin{eqnarray}
 \omega_{\mu} &=& \frac{1}{g}\pdm\varphi \quad \mbox{($O(2)$ NLSM$_2$)} \nn \\
&=& \left(\frac{2\pi}{g}\right)\frac{\sqrt{\bar{\lambda}}}{2\pi}\epsilon_{\mu\nu}\pdn\phi \quad \mbox{(SG model)} \nn \\
&=& \left(\frac{2\pi}{g}\right)\bar{\psi}\gamma_{\mu}\psi \equiv \left(\frac{2\pi}{g}\right)j_{\mu} \quad \mbox{(MT model)}.
\label{charges}
\end{eqnarray}
From Eq.\,(\ref{charges}) we can show a relation of topological
charges through the integral along the closed loop $C$,
\begin{eqnarray}
 \oint_{C}\omega_{\mu}dx^{\mu} &=& \frac{1}{g}\oint \pdm\varphi\, dx^{\mu} \,=\, \sum_{i}\,\frac{2\pi}{g}\,Q_{i}.   
\end{eqnarray}
In particular, the magnetic charge of monopoles 
(or intensity of vortices) implies 
a mass $m_{\rm \scriptscriptstyle{MT}}$ in the MT model. From the identity 
\begin{eqnarray}
& & \oint_C \omega_{\mu}dx^{\mu}\, 
= \int_S \! d^2x\,\epsilon_{\mu\nu}\,\pdm\omega_{\nu}  \nn \\
&=& \frac{2\pi}{g}\int_S\!d^2x\,\pdm(\epsilon_{\mu\nu}\bar{\psi}\gamma_{\nu}\psi) = \frac{2\pi}{g}\int_S \!d^2x\, \pdm(i\bar{\psi}\gamma_{\mu}\gamma_5\psi), 
\end{eqnarray} 
the non-zero flux of magnetic monopoles 
(non-zero topological charge) means that the
axial current $j_{5\mu} \equiv \bar{\psi}\gamma_{\mu}\gamma_5\psi$ does
not conserve classically. 
That is, a magnetic flux breaks the chiral symmetry of the
MT model at the classical level. This result agrees with the fact 
that a mass of the MT model is proportional to $\zeta$. 
If there is no topological effect, 
then $\zeta$ vanishes and a mass of the MT model becomes
zero (i.e., a massless Thirring model) where the current $j_{5\mu}$ is
conserved classically.   

Finally, we shall summarize the above results and list the relationship
between some quantities of compact QED, $O(2)$ NLSM$_2$, SG model
and MT model in Tab.\,\ref{list:tab}.

\section{Coulomb Gas Behavior and Phase Structure}

First, we shall briefly review phase structures of the
one-dimensional CG and MT model shown 
in Ref.\,\cite{steer}. In particular, a phase structure of the
MT model in the high-temperature region can be investigated 
by using exact results in a one-dimensional CG system \cite{1CG}. Next, 
we will derive the thermal pressure of the topological 
model from these results. Thus we can decide a phase structure 
in terms of a one-dimensional CG system.   
Finally, we will estimate the critical-line equation 
of the gauge theory and comment on the nature and 
order of the phase transition.

\subsection{Phase Structure of One-dimensional Coulomb Gas}
\label{CG}

A one-dimensional CG is an exactly-solvable model and some quantities
can be exactly calculated. In particular, the thermal pressure in the
one-dimensional CG system  
is very useful and powerful for investigating a phase structure 
of the MT model.  
A one-dimensional CG system, which consists of $N$ positive charged particles 
and $N$ negative ones, is described by the following partition function,  
\begin{equation}
 Z_{\rm 1CG}(z,\th,\sigma,L) \,=\, \sum_{N=0}^{\infty}\frac{z^{2N}}{(N!)^2}\,\prod_{i=1}^{2N}
\int_0^L \!\!dq_i \,\exp\left[\frac{2\pi\sigma^2}{\theta}\!\!\!\! \sum_{1 \leq i < j \leq 2N}\!\!\!\!\epsilon_i\epsilon_j\,|q_i - q_j|\right],
\end{equation}
where $\theta$ and $L$ are temperature and size of the system, respectively and $\epsilon_i =1$ for 
$i\leq N$ and $\epsilon_i = -1$ for $i \geq N+1$. The 
magnitude of the charge is denoted by $\sigma$. The fugacity $z$ 
is related to the chemical potential $\mu_{\rm \scriptscriptstyle{CG}}$
through  
\begin{equation}
 z = \sqrt{2\pi M\theta}\,\exp\left(\frac{\mu_{\rm \scriptscriptstyle{CG}}}{\theta}\right),
\end{equation}
where $M$ is a mass of particles. In the thermodynamic limit
$L\rar \infty$, the thermal pressure $P(z,\th,\sigma)$ and the mean particle density
$n(z,\th,\sigma)$ are defined by 
\begin{eqnarray}
\label{tp}
 & & P(z,\th,\sigma) \;=\; \lim_{L\rar \infty}\, \frac{\theta}{L}\ln Z_{\rm 1CG}(z,\th,\sigma,L), \\
 & & n(z,\th,\sigma) \;=\; \frac{z}{\th}\,\frac{\pd}{\pd z}P(z,\th,\sigma). 
\end{eqnarray}
The thermal pressure of a one-dimensional CG can be exactly calculated
and is given by  
\begin{equation}
 P(z,\th,\sigma) \;=\; 2\pi\sigma^2\gamma_0(\hat{z}),~~~~\hat{z}\equiv \frac{z\,\th}{2\pi\sigma^2},
\end{equation}
where $\gamma_0(\hat{z})$ is the highest eigenvalue of the Mathieu's
differential equation
\begin{equation}
\label{Mathieu}
 \left[\frac{d^2}{d\phi^2} + 2\hat{z}\cos\phi\right]\,y(\phi) = \gamma \,y(\phi)
\end{equation}
with $y(\phi + 2\pi) = y(\phi)$. The eigenvalues and eigenfunctions of
Eq.\,(\ref{Mathieu}) are well known and can be calculated by
Mathematica\footnote{We can obtain
a well-known expression of the Mathieu's differential equation 
(used in Mathematica),
\[
 y''(z) + [a - 2q\cos(2z)]y(z) = 0,
\]
through relations,
\[
 \phi = 2z,~~~\hat{z} = - \frac{q}{4},~~~\gamma = - \frac{a}{4}
\]
}.
Remark that asymptotic forms of 
$\gamma_0(\hat{z})$ for sufficiently large and small $\hat{z}$ becomes
as follows:  
\begin{eqnarray}
 & & \gamma_0(\hat{z}) \;\simeq\; 2\hat{z}^2 \,+\, O(\hat{z}^4),~~~~{\rm for}~~\hat{z} \ll 1, \\
& & \gamma_0(\hat{z}) \;\simeq\; 2\hat{z} - \sqrt{\hat{z}} \,+\, O(1),~~~{\rm for}~~\hat{z} \gg 1.
\end{eqnarray}
That is, if $\hat{z}$ is small, it behaves as a 
quadratic curve. Since $\gamma_0(\hat{z})$ is a monotonous function 
rapidly increasing like an exponential function, 
the most important parameter determining phases 
of a one-dimensional CG is $\hat{z}$, which plays a role as a certain
order parameter. 
We have summarized in Tab.\,\ref{z:tab} the relationship between the
value of $\hat{z}$, thermal pressure and phases of the CG system.
\begin{table}[htbp]
\begin{center}
{\footnotesize
 \begin{tabular}{c c c}
\hline
\qquad $\hat{z}$ \quad & \quad Pressure \quad &  \quad 1D Coulomb gas \qquad \\
\hline
\qquad Small\quad & \quad Small  \quad & \quad Molecule Phase \qquad \\
\qquad  Large \quad &  \quad Large \quad & \quad Plasma Phase \qquad  \\
\hline
 \end{tabular}
}
\end{center}
\caption{\footnotesize The parameter $\hat{z}$, thermal pressure 
and phases of the one-dimensional CG
 system.}
\label{z:tab}
\end{table}

\subsection{Phase Structure of Massive Thirring Model}
\label{CG-MT}

In the high-temperature regime, using  
the partition function of the one-dimensional CG, 
$Z_{\rm 1CG}(z,\th,\sigma,L)$, we can write  
the partition function of the MT model, $Z_{\rm MT}(T,L)$ as 
\begin{eqnarray}
\label{z-mt}
& &Z_{\rm MT}(T, L) = Z_0^F(T,L)\cdot Z_{\rm 1CG}(z,\th,\sigma,L),
\end{eqnarray}
where $z,\,\th$ and $\sigma$ are defined by
\begin{eqnarray}
& & z \,\equiv\, \frac{m_{\rm \scriptscriptstyle{MT}}^2}{2T}\left(\frac{2T}{m_{\rm \scriptscriptstyle{MT}}}\right)^{(1 + g_{\rm \scriptscriptstyle{MT}}^2/
\pi)^{-1}}\!\!\!\!\!\!\!\!\!\!\! ,\quad \th \,\equiv\, T,\quad \sigma \,
\equiv\, 
T\sqrt{\frac{\pi}{\pi + g_{\rm \scriptscriptstyle{MT}}^2}},
\end{eqnarray}
and $Z_0^F(T,L)$ is the partition function for a two-dimensional free
massless Fermi gas defined by 
\begin{equation}
 Z_0^F(T,L) \equiv \exp\left(\frac{\pi LT}{6}\right).
\end{equation}
Here we have ignored irrelevant vacuum terms. 
Parameters of a one-dimensional CG system $z,\th$ and $\sigma$ is 
expressed by those of the MT model.

\begin{figure}
 \begin{center}
  \includegraphics{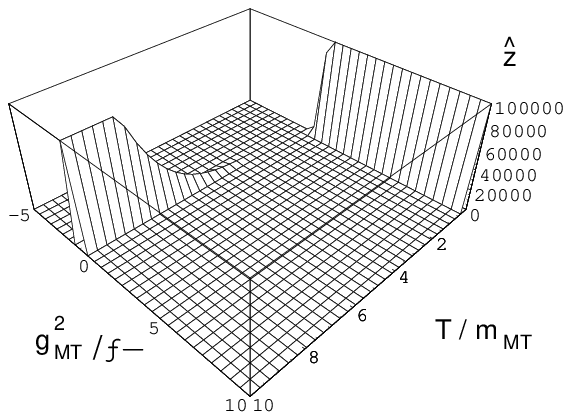}
Fig.\,\ref{z-hat:fig}\,(a)
  \includegraphics{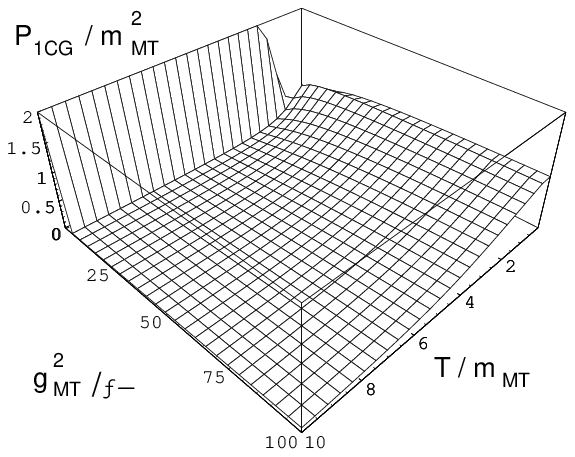}
Fig.\,\ref{z-hat:fig}\,(b)
 \end{center}
\caption{\footnotesize The behavior of $\hat{z}$ and thermal pressure
 $P_{\rm 1CG}/m^2_{\rm \scriptscriptstyle{MT}}$.}
\label{z-hat:fig}
\end{figure}

The partition function (\ref{z-mt}) leads to the thermal pressure of the MT
model, 
\begin{equation}
\label{thermal-P-MT}
 P_{\rm MT}(T) \,=\, \frac{\pi}{6}T^2 + P_{\rm 1CG}(T, g_{\rm \scriptscriptstyle{MT}}, m_{\rm \scriptscriptstyle{MT}}).
\end{equation}
The thermal pressure induced by particles forming a one-dimensional CG system,
$P_{\rm 1CG}$  
can be expressed by parameters of the MT model as 
\begin{eqnarray}
 P_{\rm 1CG}(T,g_{\rm \scriptscriptstyle{MT}}, m_{\rm \scriptscriptstyle{MT}}) &=& \frac{2\pi T^2}{1 + g_{\rm \scriptscriptstyle{MT}}^2/\pi}\, \gamma_0(\hat{z}), \nn \\
 \hat{z} &\equiv&  \frac{m_{\rm \scriptscriptstyle{MT}}^2}{4\pi T^2} \left(1 + \frac{g_{\rm \scriptscriptstyle{MT}}^2}{\pi}\right)\left(\frac{2T}{m_{\rm \scriptscriptstyle{MT}}}\right)^{(1 + g_{\rm \scriptscriptstyle{MT}}^2/\pi)^{-1}}.
\end{eqnarray}
The numerical plot of $\hat{z}$ as a function of $T/m_{\rm \scriptscriptstyle{MT}}$
and $g_{\rm \scriptscriptstyle{MT}}^2/\pi$ is shown in Fig.\,\ref{z-hat:fig}. 
A cliff and slope appear in Fig.\,\ref{z-hat:fig}\,(a). 
These  mean that the thermal pressure is very strong,  
and hence a one-dimensional CG system is in a plasma phase. 
There exists a slope in the negative-coupling and high-temperature 
regime but we cannot believe this result since it is out of the DR
regime \cite{steer},  
\begin{eqnarray}
& & T \gg m_{\rm \scriptscriptstyle{MT}},\quad g_{\rm \scriptscriptstyle{MT}}^2 > 0, \\
& & T \gtrsim m_{\rm \scriptscriptstyle{MT}},\quad g_{\rm \scriptscriptstyle{MT}}^2/\pi \gg 1,
\end{eqnarray}
where the expression of the thermal pressure (\ref{thermal-P-MT}) 
is reliable. The
constraints for the temperature imply the region where the dimensional
reduction (a one-dimensional CG approximation) is valid. Those for the
coupling ensure the existence of the thermo-dynamical limit.       

It should be remarked that the first term in Eq.\,(\ref{thermal-P-MT}) 
implies the finite-size effect of the cylinder (i.e., geometrical force) 
as discussed in Appendix \ref{pressure:app}, 
and hence 
it does not depend on parameters of the MT model and universal. 
Thus we will omit this term in studying the phase
structure since it is purely geometrical contribution.  

In the three specific regions of the MT model, the behavior of a 
one-dimensional CG system  
has been analytically 
investigated \cite{steer}, and 
we shall list results of Ref.\,\cite{steer} below 
for the purpose of studying compact QED.
\begin{enumerate}
\renewcommand{\labelenumi}{\Roman{enumi}.}
 \item~~ $g_{\rm \scriptscriptstyle{MT}}^2 \,\gg\, \pi$ and $T\, \gg\, m_{\rm \scriptscriptstyle{MT}}g_{\rm \scriptscriptstyle{MT}}/2\pi$:

Then, $\hat{z} \ll 1$, and the thermal pressure becomes as 
\begin{equation}
\frac{P_{\rm 1CG}}{m^2_{\rm \scriptscriptstyle{MT}}} \; \sim\; 
\left(\frac{m_{\rm \scriptscriptstyle{MT}}g_{\rm \scriptscriptstyle{MT}}}{2\pi T}\right)^2\; \ll\; 1.
\end{equation}
Thus the thermal pressure vanishes as $T\rar \infty$ and hence the 
system is in the ``molecule'' phase. 
 \item~~ $g_{\rm \scriptscriptstyle{MT}}^2\, \gg\, \pi$ and $T\, \ll\, m_{\rm \scriptscriptstyle{MT}}g_{\rm \scriptscriptstyle{MT}}/2\pi$:
 
In this region, $\hat{z} \gg 1$ and the thermal pressure behaves as 
\begin{equation}
\frac{P_{\rm 1CG}}{m_{\rm \scriptscriptstyle{MT}}^2} \;\sim\; 1 - 
\frac{\pi T}{m_{\rm \scriptscriptstyle{MT}}g_{\rm \scriptscriptstyle{MT}}} \;\gg\; \frac{1}{2}.
\end{equation}
Thus, the system is in the ``plasma'' phase.
 \item~~ $0\, < \,g_{\rm \scriptscriptstyle{MT}}^2 \,\ll\, 1$ and
       $T \,\gg\, m_{\rm \scriptscriptstyle{MT}}/2\pi$: 

We have  $\hat{z} \,\simeq\, m_{\rm \scriptscriptstyle{MT}}/2\pi T$,
      which leads  to $\hat{z}\ll 1$. Thus the system is in the 
``molecule'' phase and the thermal pressure behaves as 
\begin{equation}
\frac{P_{\rm 1CG}}{m^2_{\rm \scriptscriptstyle{MT}}} \;\sim\; \frac{1}{\pi}.
\end{equation}
\end{enumerate}

\subsection{Phase Structure of Compact QED}

\begin{figure}
 \begin{center}
  \includegraphics[scale=.8]{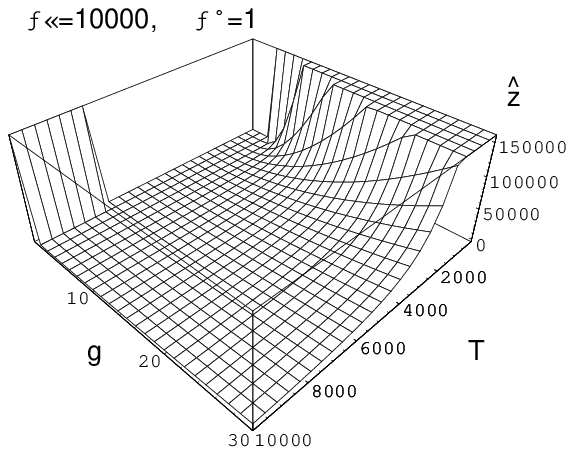}
Fig.\,\ref{abc:fig}\,(a)
  \includegraphics[scale=.8]{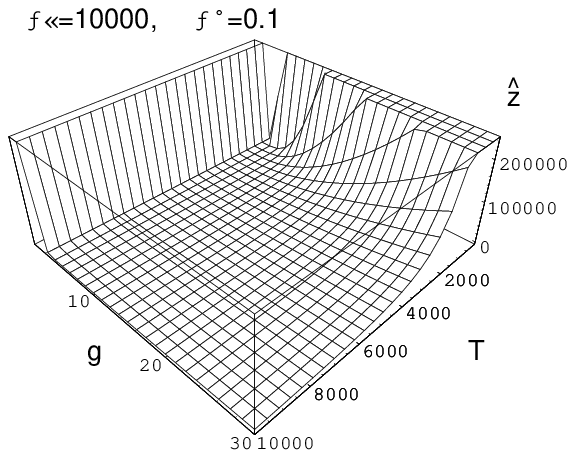}
Fig.\,\ref{abc:fig}\,(b)
  \includegraphics[scale=.8]{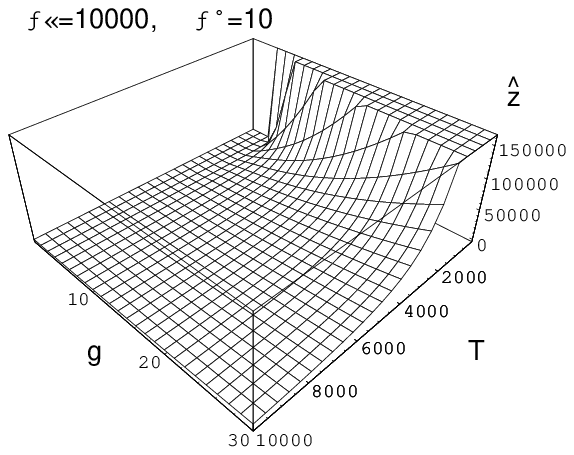}
Fig.\,\ref{abc:fig}\,(c)
  \includegraphics[scale=.8]{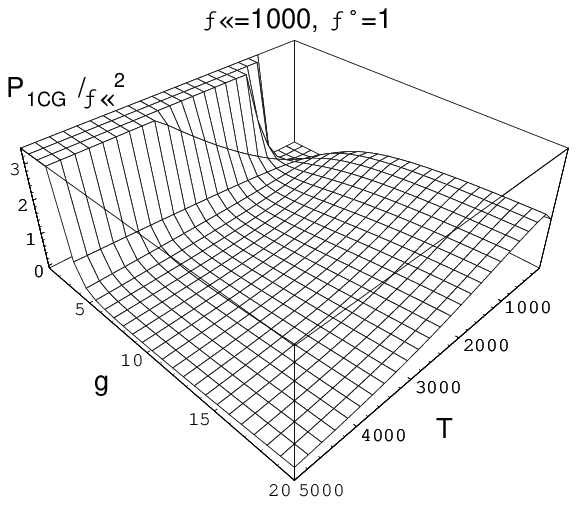}
Fig.\,\ref{abc:fig}\,(d)
 \end{center}
\caption{\footnotesize The plot of $\hat{z}$ is shown in
 Figs.\,\ref{abc:fig}\,(a),\,(b) and (c) at fixed $\Lambda =10000$, and
 $\kappa = 1,\,0.1$ and $10$, respectively. In Fig.\,\ref{abc:fig}\,(d),
 the dimensionless quantity $P_{\rm 1CG}/\Lambda^2$ is plotted.}
\label{abc:fig}
\end{figure}

We can translate previous results of the CG system (MT model) 
into compact QED.
By the use of Eq.\,(\ref{MT-gauge}), we
can rewrite the thermal pressure and its argument 
of the one-dimensional CG system (MT model)   
in terms of compact QED as follows: 
\begin{eqnarray}
& & P_{\rm 1CG}(T,g,\Lambda,\kappa) \;=\; \frac{4\pi^3T^2}{g^2}\,\gamma_0(\hat{z}), \\
& & \hat{z} \;=\;
\frac{g^2}{2\pi^3}\left(\frac{2T^2}{\zeta\Lambda^2}\right)^{(\pi^2 - g^2)/g^2} \,=\, \frac{g^2}{2\pi^3}\left(\frac{2 T^2}{\Lambda^2}\right)^{(\pi^2 - g^2)/g^2}
\kappa^{2\pi^2(g^2 - \pi^2)/g^4}.
\end{eqnarray}
We have numerically plotted $\hat{z}$ and thermal pressure $P_{\rm
1CG}/\Lambda^2$ as functions of $T$ and $g$
at fixed $\Lambda$ and $\kappa$ in Fig.\,\ref{abc:fig}.  
It can be easily seen that the theory with strong coupling is confined. 
That is, a confining phase similar to the lattice gauge theory 
appears in the strong-coupling regime.    
Non-trivial phase structure tends to disappear as $\kappa$ becomes small. 

Here, we should remember the validity of our calculations.  
Recall that we have used the one-dimensional CG approximation and 
equivalence between a one-dimensional CG
and high-temperature MT model in our derivation. The validity of the 
approximation and equivalence is restricted within 
the DR regime where the existence of the thermo-dynamical
limit is also guaranteed. This regime of compact QED is expressed by 
\begin{eqnarray}
& & T \gg \sqrt{2}\,\kappa^{\pi^2/g^2}\Lambda,\quad g^2 > 2\pi^2, \\
& & T \gtrsim \sqrt{2}\,\kappa^{\pi^2/g^2}\Lambda,\quad g^2 \gg 4\pi^2.
\end{eqnarray}
The constraints for the coupling are obstacles to propose 
the existence of another confining phase. 
As discussed in Ref.\,\cite{steer}, we might formally
take the gauge-coupling $g$ arbitrary from the viewpoint of the CG
system, although extra renormalizations at least would be required from the
standpoint of the MT model. Also, compact QED would possibly 
have the same renormalization problem as in the MT model. 
In addition, it is quite well known that the
density of magnetic monopoles decreases rapidly as the coupling
constant gets smaller \cite{BMK}. That is, monopole effects would almost
vanish in the weak-coupling region. Thus we conclude and strongly remark 
that the confinement at 
weak-coupling and high-temperature would not appear.

\begin{figure}
 \begin{center}
  \includegraphics{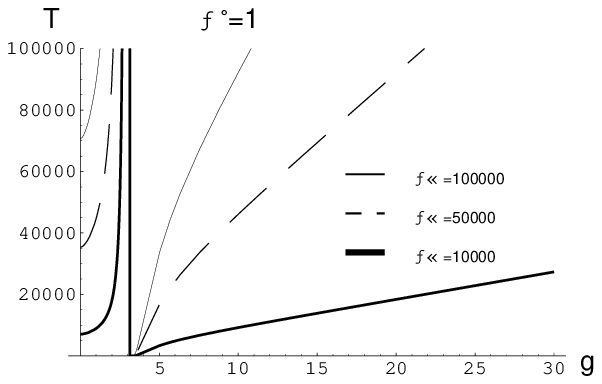}
Fig.\,\ref{cri:fig}\,(a)
  \includegraphics{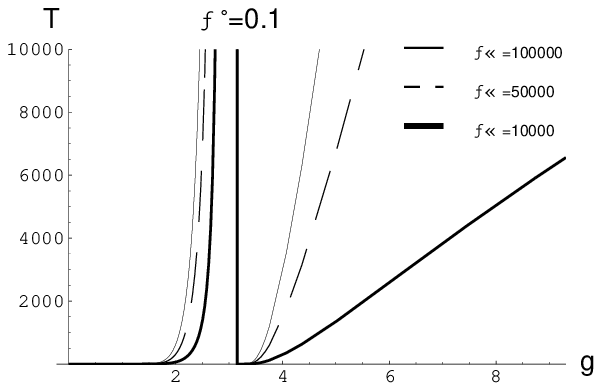}
Fig.\,\ref{cri:fig}\,(b)
  \includegraphics{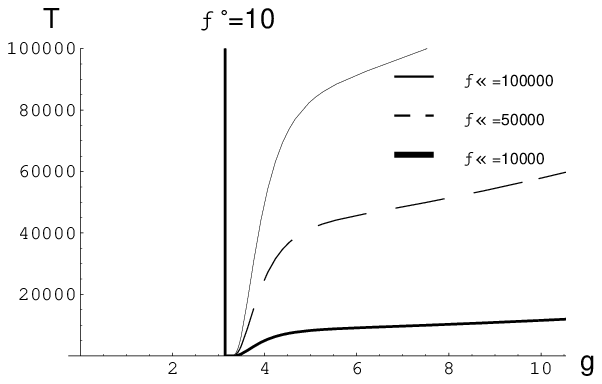}
Fig.\,\ref{cri:fig}\,(c)
 \end{center}
\caption{\footnotesize Critical-line equations: 
In Figs. \ref{cri:fig}\,(a),\,(b) and (c) 
the critical-lines for $l=1$ are plotted at fixed values.}
\label{cri:fig}
\end{figure}

We can also estimate the critical temperature as a function of $g$ at
 fixed values of $\Lambda,\,\mu$ and $\kappa$  
by setting $\hat{z} \equiv l ~({\rm const})$. 
The critical-line equation can be evaluated by 
\begin{eqnarray}
T_{\rm cr} &=& \frac{\zeta^{1/2}\Lambda}{\sqrt{2}}\left(\frac{2\pi^3 l}{g^2}\right)^{g^2/2(\pi^2 - g^2)} \nn \\
&=& \frac{\Lambda}{\sqrt{2}} \,\kappa^{\pi^2/g^2} 
\left( \frac{2\pi^3 l }{g^2}\right)^{g^2/2(\pi^2 - g^2)}.
\label{critical}
\end{eqnarray}
The critical temperature (\ref{critical}) is numerically plotted 
in Fig.\,\ref{cri:fig}.
In particular, in the strong-coupling region, 
we can obtain from Eq.\,(\ref{critical}) the asymptotic line,   
\begin{eqnarray}
\label{asymp}
 T_{\rm cr}^{\rm asy} \;\simeq\; \frac{\Lambda\,\kappa^{\pi^2/g^2 }}{2\pi^{3/2}l^{1/2}}\,g,
\end{eqnarray}
where if we take $l = 1/16$, Eq.\,(\ref{asymp}) is precisely 
identical with 
the asymptotic line given by a {\it perturbative} method 
in our previous work \cite{KY3}. A one-dimensional CG approximation does
not depend on a perturbation theory in the two-dimensional theory, 
and hence this result is
non-trivial and further confirms the validity of our previous result for
the asymptotics of the critical-line equation 
at sufficiently high-temperature and strong-coupling 
region of compact QED. 

\begin{figure}
 \begin{center}
  \includegraphics[scale=.8]{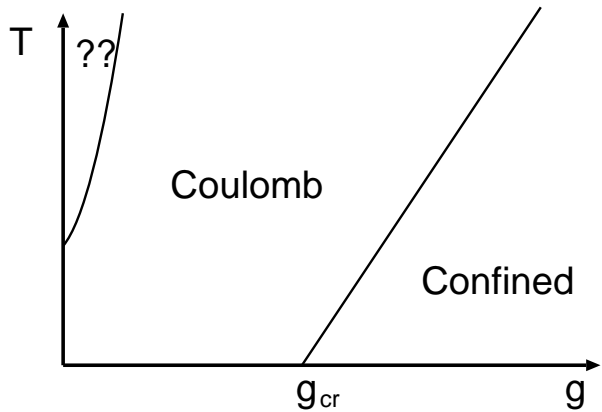}
 \end{center}
\caption{\footnotesize Phase diagram of a compact $U(1)$ gauge theory.}
\label{phase-gauge:fig}
\end{figure}

In conclusion, the phase structure shown in
Fig.\,\ref{phase-gauge:fig} is obtained, and 
agrees with our previous results \cite{KY2,KY3}.

Moreover, we can analytically investigate the specific three regime  
denoted in subsection \ref{CG-MT}. The results are listed below. 
\begin{enumerate}
\renewcommand{\labelenumi}{\Roman{enumi}.}
 \item~~$g^2\, \gg\, 4\pi^2,~~ T\, \gg\, 
T_{{\rm cr}\,(l=1)}^{\rm asy}\, $: 

In this region, $\hat{z}\,\ll\, 1$ and 
\[
P_{\rm 1CG} \;\sim\; \frac{\Lambda^4g^2}{2\pi^3T^2}\,\kappa^{4\pi^2/g^2} \!\gg\, 2\Lambda^2\,\kappa^{2\pi^2/g^2}.
\] Thus a one-dimensional CG is in the ``molecule'' phase and hence
 compact QED is in a deconfining phase (Coulomb phase). 
 \item ~~$g^2 \,\gg\, 4\pi^2,~~ T\, \ll\,T_{{\rm cr}\,(l=1)}^{\rm asy}    \,$: 

Now $\hat{z}\,\gg\, 1$ and 
\[
P_{\rm 1CG} \;\sim\; 2\Lambda^2\,\kappa^{2\pi^2/g^2} \! -\, 
\frac{2\pi^{3/2}\Lambda T}{g}\,\kappa^{\pi^2/g^2} \!\ll\, \Lambda^2\,\kappa^{2\pi^2/g^2}.
\] Thus a one-dimensional CG is 
in the ``plasma'' phase, and so the gauge theory is confined. 
Of course, this result also agrees with our previous result.  
 \item~~$2\pi^2 \, < \, g^2 \ll 2\pi^2 + 2\pi,~~T \,\gg\, (\Lambda/\sqrt{2}\pi)\,\kappa^{\pi^2/g^2}\,  $: 

In this region, $\hat{z} \,\ll\, 1$ and 
\[
P_{\rm 1CG} \;\sim\; \frac{2\Lambda^2}{\pi}\,\kappa^{2\pi^2/g^2}.
\] Thus a one-dimensional CG is in the ``molecule'' phase, and hence the gauge theory should be in the Coulomb phase. This result has also no contradiction. 
\end{enumerate}

It should be remarked that the critical-line equation derived in 
our previous work \cite{KY3} appears in the region I, II.

\paragraph{Comments on the Order of the Confining Transition}

\begin{figure}
 \begin{center}
  \includegraphics[scale=.8]{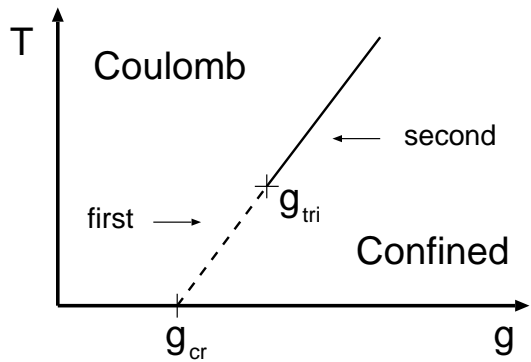}
 \end{center} 
\caption{\footnotesize The order of the confining-deconfining phase
 transition. If the order is first at zero temperature, then the
 tri-critical point exists.  }
\label{phase:fig}
\end{figure}

In general, a photon mass is taken as the order-parameter \cite{wilson}
of the confinement in compact QED.   
The confining transition of the
four-dimensional compact QED is 
second-order if it is second-order at zero temperature \cite{wilson,Peskin}. 
If it is first-order at zero temperature, then the tri-critical
point exists on the critical-line as shown in Fig.\,\ref{phase:fig}.

Several numerical simulation results of the compact $U(1)$ lattice 
gauge theory have been reported until now.  
However, some of them claim that it is first-order,
and others show that it is second. It seems rather difficult to
decide the order numerically in the lattice formulation. It is because
we have a continuum-limit problem as noted in Introduction. 
Our scenario is based on the continuum formulation and it would be
possible to avoid such a problem. However, in our scenario, we can only 
guess the nature of the phase transition is BKT-type and still have no
clear answer. The photon mass cannot be calculated yet. 
This is one of the most interesting future problems.

\section{Fermion Condensate versus Monopole Condensate}

In this section we will 
discuss the correspondence between fermion condensate in the MT model and
monopole condensate in compact QED. In particular, the chiral
operators of the MT model correspond to the monopoles and anti-monopoles
in compact QED. Again, the exact results of 
the one-dimensional CG system are utilized.  
It can be also seen that the confining phase transition in compact QED
is closely related to the chiral symmetry restoration in the MT model. 
This relation seems 
natural from the viewpoint of the perturbed CFT.

\subsection{Fermion Condensate and Chiral Symmetry in Massive
  Thirring  Model}

We will review the equivalence between the chiral symmetry
restoration in the MT model and the behavior of the 
one-dimensional CG system \cite{steer}. 
The charged particles consisting of the CG system 
could be interpreted as monopoles and anti-monopoles in compact QED. 
Thus the
chiral symmetry in the MT model is intimately related to the behavior of
monopoles in compact QED. In particular, 
the fermion condensate corresponds to the monopole 
condensate.  

In the DR regime, utilizing the exact results for the one-dimensional CG
system, the fermion condensate in the MT model can be calculated as 
\begin{equation}
\label{thermal-MT}
 \la\la\bar{\psi}\psi \ra\ra \;=\; \frac{\partial}{\partial\, m_{\rm MT}}P_{\rm MT}(T).
\end{equation}
We now consider the system coupled with the thermal bath 
and so the vacuum expectation value is replaced with the thermal
expectation value $\la\la \;\ldots\; \ra\ra$ defined by 
\begin{equation}
 \la\la\;\ldots\;\ra\ra \equiv \frac{1}{Z(T)}{\rm Tr}\left(\;\ldots\; \e^{-\beta \hat{H}}\right) \,=\, \frac{N_{\beta}}{Z(T)}\int_{\rm anti-periodic} \!\!\!\!\! \!\!\!\!\!\!\!\!\!\! \!\!\!\!\! [d\psi][d\bar{\psi}]\;\ldots\;\exp(-S[\psi,\bar{\psi}]),
\end{equation}
where $N_{\beta}$ is an infinite $T$-dependent constant arising in the
path-integral description. 

Here, we shall introduce the
one-charge density functions $f_{\pm}(z,\th,\sigma,L;X)$ defined by 
\begin{eqnarray}
 & & f_+(z,\th,\sigma,L;X)  \nn \\ 
&=& \frac{1}{Z_{\rm 1CG}}\sum_{N=1}^{\infty}\frac{z^{2N}}{(N!)^2}\left(\prod_{i=1}^{2N}\int_0^{L}\!\!dx_i\right)\sum_{j=1}^{N}\del(x_j - X) \exp\left[\frac{2\pi\sigma^2}{\th}\!\!\!\!\!\sum_{1\leq j < i \leq 2N}\!\!\!\!\epsilon_i\epsilon_j|x_i - x_j|\right] \nn \\
&=& \frac{1}{Z_{\rm 1CG}}\sum_{N=1}^{\infty}\frac{z^{2N}}{N!(N-1)!}\left(\prod_{i=1}^{2N}\int_0^{L}\!\!dx_i\right)\del(x_N - X) \exp\left[\frac{2\pi\sigma^2}{\th}\!\!\!\!\!
\sum_{1\leq j < i \leq 2N}\!\!\!\!\epsilon_i\epsilon_j|x_i - x_j|\right].
\nn \\
\label{f+-}
\end{eqnarray}
Note that the translational invariance is
restored in the limit $L\rar\infty$ since a simple shift $x_i\rar x_i +
X,\,(^{\forall}i=1,\ldots,2N)$ ensures that $f_{+}$ is independent of
$X$. The $f_+$ and $f_-$ correspond to the positive charged particle and
negative one, respectively. Those also obey the relation $f_+ =
f_-$. 
By the use of the following identity 
\begin{equation}
 f_+(z,\th,\sigma,L) \;\equiv\;\frac{1}{L}\int^L_0 \! dX\,f_+ 
(z,\th,\sigma,L;X) = \frac{1}{2L}\,z\,\frac{\partial}{\partial z}\ln Z_{\rm 1CG}, 
\end{equation}
we obtain the expression 
\begin{equation}
 m_{\rm \scriptscriptstyle{MT}}\la\la\bar{\psi}\psi\ra\ra \;=\; 2T f_+\left(
z=\frac{m_{\rm \scriptscriptstyle{MT}}^2}{2T}\left(\frac{2T}{m_{\scriptscriptstyle{MT}}}\right)^{(1 + g^2_{\rm \scriptscriptstyle{MT}}/\pi)^{-1}},\th=T,\sigma=T\sqrt{\frac{\pi}{\pi + g_{\rm \scriptscriptstyle{MT}}^2}},L\right).
\label{1-point-MT}
\end{equation}
The relation (\ref{1-point-MT}) suggests 
that the fermion condensate in the MT model is 
described by the particle density in the CG system. 
Finally, the following expression of the fermion condensate 
in the MT model as 
\begin{equation}
\label{fermion-condensate}
 \la\la\bar{\psi}\psi \ra\ra \;=\; \frac{m_{\rm \scriptscriptstyle{MT}}}{2}\left(\frac{2T}{m_{\rm \scriptscriptstyle{MT}}}
\right)^{(1 + g_{\rm \scriptscriptstyle{MT}}^2/\pi)^{-1}}\!\!
\gamma_0' \,\left[\,\frac{m_{\rm \scriptscriptstyle{MT}}^2}{4\pi T^2}\left(1 + \frac{g_{\rm \scriptscriptstyle{MT}}^2}{\pi}\right) 
\left(\frac{2T}{m_{\rm \scriptscriptstyle{MT}}}\right)^{
(1+g_{\rm \scriptscriptstyle{MT}}^2/\pi)^{-1}} \right].
\end{equation}
\begin{figure}
 \begin{center}
  \includegraphics{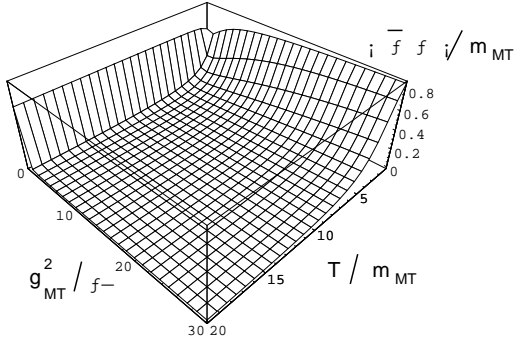}
 \end{center}
\caption{\footnotesize Fermion condensate in the MT model. }
\label{fermion:fig}
\end{figure}
The numerical plot of the fermion condensate (\ref{fermion-condensate})
is shown in Fig.\,\ref{fermion:fig}. The condensate becomes smaller as
$T$ grows at strong coupling. It should be noted again that 
the validity of the fermion condensate (\ref{fermion-condensate}) 
is restricted within the DR regime. Thus, the condensate grows 
as $T$ becomes large in the negative-coupling regime but 
we cannot believe this result since it is out of the DR regime. 

In particular, we list the result in three regions below.  
\begin{enumerate}
\renewcommand{\labelenumi}{\Roman{enumi}.}
 \item~~$g_{\rm \scriptscriptstyle{MT}}^2\, \gg\, \pi,~~T\,\gg\, m_{\rm \scriptscriptstyle{MT}}g_{\rm \scriptscriptstyle{MT}}/2\pi$\,: 

We obtain 
\[
\la\la \bar{\psi}\psi \ra\ra\; \sim\; 
 2m_{\rm \scriptscriptstyle{MT}}\left(\frac{m_{\rm \scriptscriptstyle{MT}}g_{\rm \scriptscriptstyle{MT}}}{2\pi T} \right)^2 \, \ll\, 2m_{\rm 
\scriptscriptstyle{MT}}.
\]
In the limit $T\rar\infty$, The condensate vanishes asymptotically and the chiral symmetry breaking is restored.
 \item~~$g^2_{\rm \scriptscriptstyle{MT}} \,\gg\, \pi,~~T\,\ll\, m_{\rm \scriptscriptstyle{MT}}g_{\rm \scriptscriptstyle{MT}}/2\pi$\,: 

The fermion condensate is approximately  
\[
\la\la \bar{\psi}\psi\ra\ra \;\sim\; m_{\rm \scriptscriptstyle{MT}}\left(
\frac{2T}{m_{\rm \scriptscriptstyle{MT}}} \right)^{\pi/g_{\rm \scriptscriptstyle{MT}}^2}.
\] If $T$ is low enough, the condensate is also small. 
Note that the above result is valid only in the restricted region, $T\, \gtrsim\, m_{\rm \scriptscriptstyle{MT}}$. 
 \item~~$0 \,< \, g_{\rm \scriptscriptstyle{MT}}^2 \,\ll\, 1,~~T\, \gg\, m_{\rm \scriptscriptstyle{MT}}/2\pi$\,: 

The condensate tends to the constant value 
\[
\la\la\bar{\psi}\psi \ra\ra \;\sim\; \frac{2 m_{\rm \scriptscriptstyle{MT}}}{\pi}.
\]
\end{enumerate}

Moreover, we can define the parameter $q_0$ denoting the mean
inter-particle distance by the inverse of the charge density, 
$q_0 = \frac{1}{f_+}$, 
regardless of whether those are positively or negatively charged. 
Thus it is expressed by 
\begin{eqnarray}
 q_0(g_{\rm \scriptscriptstyle{MT}}, T) & = & \frac{2T}{m_{\rm\scriptscriptstyle{MT}} \la\la\,\bar{\psi}\psi\,\ra\ra } \nn \\ 
&=& T \left(\frac{m_{\rm \scriptscriptstyle{MT}}}{2T}
\right)^{(1 + g_{\rm \scriptscriptstyle{MT}}^2/\pi)^{-1}}\!\!
\gamma_0' \,\left(\hat{z}\right)^{-1}.
\end{eqnarray}

\paragraph*{Chiral Operators in Massive Thirring Model}
The chiral operators (fermion density) of the MT model 
\begin{equation}
 \sigma_{\pm} \;=\; \bar{\psi}\left(\frac{1\pm\gamma^5}{2}\right)\psi,
\end{equation}
corresponds to the positive charged particle and negative one,
respectively. In terms of the components $\psi_1$ and $\psi_2$,
$\sigma_{\pm}$ are defined by $\sigma_+\,=\, \psi^{\dagger}_2\psi_1$ and
$\sigma_- \,=\, \psi^{\dagger}_1\psi_2 $.
The $\sigma_{\pm}$ transforms under the chiral
transformation $\psi \rar \e^{ia\gamma^5}\psi$ as 
\begin{equation}
\label{chiral-charge}
 \sigma_{\pm} \longrightarrow \exp(\pm 2i a) \sigma_{\pm}.
\end{equation}
In other words, Eq.\,(\ref{chiral-charge}) implies that 
$\sigma_{\pm}$ has a well-defined $\pm$ chiral charge.   
Thus the chiral invariant combination $\sigma_{+}\sigma_-$ is
neutral in terms of the chiral charge. This neutral pair corresponds to
the ``molecule'' in the CG system.

\subsection{Monopole Condensate in Compact QED}

We can interpret fundamental fermions in the MT model as monopoles in
compact QED and so treat explicitly the monopole dynamics 
from that of the MT model. Recall that 
charged particles consisting of the CG system 
are inherently related to monopoles in compact QED. 
The positively and negatively charged particles in the CG system 
correspond to the chiral operators $\sigma_+$ and $\sigma_-$,
respectively. In the four-dimensional language, a monopole and
an anti-monopole correspond to the chiral operator $\sigma_+$ and
$\sigma_-$, respectively. It would be expected that 
we could define the monopole (anti-monopole) operator $M$ ($\bar{M}$)  
by the chiral operator $\sigma_+$ ($\sigma_-$). 
Roughly speaking, we can effectively treat the quantum theory of
monopoles as the chiral operators in the two-dimensional MT model
through the PS mechanism, and 
the quantum MT model in two dimensions describes 
the effective theory of monopoles in four-dimensional compact QED.  

Thus we can identify the monopole condensate
in compact QED with the fermion condensate in the MT model as follows: 
\begin{equation}
 \la\la\,M \,\ra\ra \;\equiv\; \la\la\,\bar{\psi}\psi\,\ra\ra,
\end{equation}  
where $\la\la\,\bar{M} \,\ra\ra \,=\, \la\la\,M\,\ra\ra$ since the net
monopole charge is zero.

We can express the fermion condensate by parameters of compact QED 
as follows:
\begin{eqnarray}
\label{monopole}
 \la\la\,M\, \ra\ra &=& \frac{\Lambda}{\sqrt{2}}
\left(\frac{\sqrt{2}T}{\Lambda}\right)^{2\pi^2/g^2}\kappa^{\pi^2(g^2-2\pi^2)/g^4}\gamma_0'(\hat{z}), \\
\hat{z} &=& \frac{g^2}{2\pi^3}\left(\frac{2T^2}{\Lambda^2}\right)^{(\pi^2-g^2)/g^2}\kappa^{2\pi^2(g^2-\pi^2)/g^4}.
\end{eqnarray}
\begin{figure}
 \begin{center}
  \includegraphics{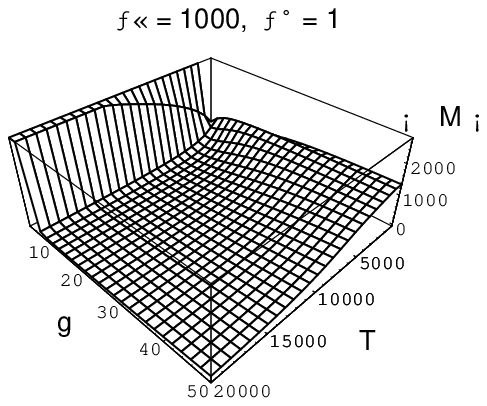}
Fig.\,\ref{monopole:fig}\,(a)
  \includegraphics{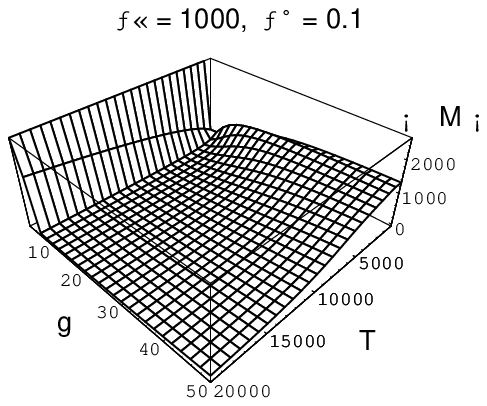}
Fig.\,\ref{monopole:fig}\,(b)
  \includegraphics{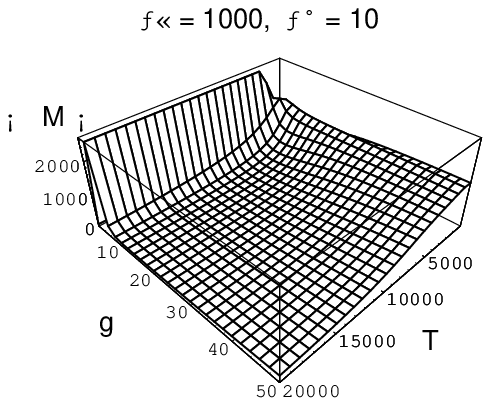}
Fig.\,\ref{monopole:fig}\,(c)
  \includegraphics{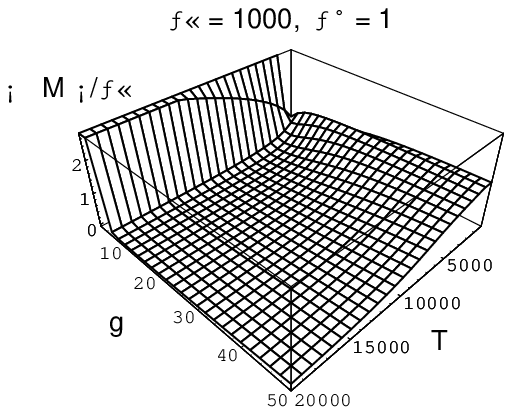}
Fig.\,\ref{monopole:fig}\,(d)
 \end{center}
\caption{\footnotesize Monopole Condensate.}
\label{monopole:fig}
\end{figure}
The numerical plot of Eq.\,(\ref{monopole}) is shown in
Fig.\,\ref{monopole:fig}. The behavior of the monopole condensate is
almost same as the fermion condensate in the MT model. The condensate
vanishes as $T\rar \infty$ at strong coupling. In the weak-coupling
regime, the condensate also 
becomes large as the temperature grows but this result
is doubtful since it is out of the DR regime. 
In the same way as the MT model, we would not be able to 
believe the results at weak-coupling. 

Finally, we shall investigate analytically the condensate in the specific 
three regimes and list the results below.  
\begin{enumerate}
\renewcommand{\labelenumi}{\Roman{enumi}.}
 \item~~$g^2 \,\gg\, 4\pi^2,~~ T \,\gg\, T^{\rm asy}_{{\rm cr}\,(l=1)}$\,: 

 The condensate is approximately 
\[
\la\la\,M\, \ra\ra \;\sim\; 
\frac{\Lambda\, g^2}{\sqrt{2}\pi^3}\left(\frac{\Lambda}{T}\right)^2\,\kappa^{3\pi^2/g^2} \!\ll \, 2\sqrt{2}\Lambda\,\kappa^{\pi^2/g^2}.
\]
 \item~~$g^2 \,\gg\, 4\pi^2,~~ T\, \ll\,T^{\rm asy}_{{\rm cr}\,(l=1)} $\,: 

The condensate becomes 
\[
\la\la\,M\, \ra\ra \;\sim\; 
\sqrt{2}\Lambda\,\kappa^{\pi^2/g^2}\left(\frac{\sqrt{2} T}{\kappa^{\pi^2/g^2}\Lambda}\right)^{\pi^2/g^2}.
\]

 \item~~$2\pi^2 \,<\, g^2 \,\ll\, 2\pi^2 + 2\pi,~~T \,\gg\, (\Lambda/\sqrt{2}\pi)\,\kappa^{\pi^2/g^2}$\,:

 Now, we obtain the condensate 
\[
\la\la \,M\, \ra\ra \;\sim\; \frac{2\sqrt{2}\Lambda}{\pi}\,\kappa^{\pi^2/g^2}.
\] 
\end{enumerate}
\begin{table}
\begin{center}
{\footnotesize
 \begin{tabular}{|c|c|c|c|}
\hline
  & MT at high temperature & One-dimensional CG & compact QED \\
\hline \hline
Thermal Pressure & Fermions  & Vortices  & Monopoles   \\
\hline
Large & Chirally Broken  & Plasma   & Confinement \\
Small & Chirally Symmetric & Molecule & Coulomb \\
\hline
\hline
Condensate & Fermion Condensate & Particle Density & Monopole Condensate
  \\
\hline
Non-zero & Chirally Broken & Plasma & Confinement \\
Zero & Chirally Symmetric & Molecule  & Coulomb \\
\hline
\hline
Operators & $\sigma_+,\;\sigma_-$,\,(chiral operators) & Vortex,\,Anti-vortex &
  Monopole,\,Anti-monopole \\
\hline
1-point Function & $\la\la\,\bar{\psi}\psi\,\ra\ra$\;$\left(\,\la\la\, \sigma_+\,\ra\ra\,\right)$ & $f_+ \;(\,=\,f_-\,)$  & $\la\la\, M\, \ra\ra \; (\,=\,\la\la\, \bar{M}\,\ra\ra\,)$  \\
2-point Function &   $\la\la\, \sigma_+\sigma_-\, \ra\ra$,\;$\la\la\, \sigma_+\sigma_+\, 
\ra\ra$ 
& $f_{+-},\;f_{++}$  & $\la\la\, M\bar{M}\,\ra\ra$,\;$\la\la\, MM \,
\ra\ra$ \\
\hline
\end{tabular}
}
\end{center}
\caption{\footnotesize Relations of observables and phases in 
the MT model at high temperature, one-dimensional CG system and compact QED. }
\label{tab2:fig}
\end{table}

The mean inter-monopole distance can be rewritten by 
\begin{eqnarray}
 q_0(g_{\rm \scriptscriptstyle{MT}}, T) & = & 
\frac{1}{\la\la \, M \bar{M} \,\ra\ra } \sqrt{\frac{2}{\zeta}}
\frac{T}{\Lambda} \nn \\ 
&=& T \left(\sqrt{\frac{\zeta}{2}}\frac{\Lambda}{T}
\right)^{2\pi^2/g^2}
\gamma_0' \,\left(\hat{z}\right)^{-1}.
\end{eqnarray}

\subsection{Two-charge Correlators}

The two-charge density functions in the CG system is given 
in Ref.\,\cite{1CG} by
\begin{eqnarray}
f_{+-}(X,\,Y) 
&=& \frac{1}{Z_{\rm 1CG}}\sum_{N=1}^{\infty}\frac{z^{2N}}{(N!)^2}\left(\prod_{i=1}^{2N}\int^L_0\!\!dx_i\right)\sum_{i=1}^{N}\sum_{j=N+1}^{2N}\del(x_i-X)\del(x_j-Y) \nn \\
& & \times \exp\left[\frac{2\pi\sigma^2}{\th}\!\!\!\!\sum_{1\leq i < j \leq 2N}\!\!\!\epsilon_i\epsilon_j|x_i-x_j|\right] \nn \\
&=& \frac{1}{Z_{\rm 1CG}}\sum_{N=1}^{\infty}\frac{z^{2N}}{((N-1)!)^2}\left(\prod_{i=1}^{2N}\int^L_0\!\!dx_i\right)\del(x_N-X)\del(x_{2N}-Y) \nn \\
& & \times\exp\left[\frac{2\pi\sigma^2}{\th}\!\!\!\!\sum_{1\leq i < j \leq 2N}\!\!\!\epsilon_i\epsilon_j|x_i-x_j|\right], \label{2-point-1}
\\
f_{++}(X,\,Y) 
&=& \frac{1}{Z_{\rm 1CG}}\sum_{N=1}^{\infty}\frac{z^{2N}}{(N!)^2}\left(\prod_{i=1}^{2N}\int^L_0\!\!dx_i\right)\sum_{i=1}^{N}\sum_{i \neq j= 1}^{N}\del(x_i-X)\del(x_j-Y) \nn \\
& & \times \exp\left[\frac{2\pi\sigma^2}{\th}\!\!\!\!\sum_{1\leq i < j \leq 2N}\!\!\!\epsilon_i\epsilon_j|x_i-x_j|\right] \nn \\
&=& \frac{1}{Z_{\rm 1CG}}\sum_{N=1}^{\infty}\frac{z^{2N}}{N!(N-2)!}\left(\prod_{i=1}^{2N}\int^L_0\!\!dx_i\right)\del(x_{N-1})\del(x_{N}-Y) \nn \\
& & \times\exp\left[\frac{2\pi\sigma^2}{\th}\!\!\!\!\sum_{1\leq i < j \leq 2N}\!\!\!\epsilon_i\epsilon_j|x_i-x_j|\right].
\label{2-point-2}
\end{eqnarray}
By the use of Eq.\,(\ref{2-point-1}) and (\ref{2-point-2}), we can
write the correlators of the chiral operators in the MT model 
at finite temperature, $\la\la\,\sigma_+(X)\sigma_-(0)\,
\ra\ra$ and $\la\la\,\sigma_+(X)\sigma_+(0)\,\ra\ra$ as 
\begin{eqnarray}
& & \la\la\,T_C\sigma_+(X)\sigma_-(0) \,\ra\ra \nn \\
&=& \left(\frac{T}{m_{\rm \scriptscriptstyle{MT}}}\right)^2
\frac{1}{\left[4Q^2(X_1,X_2)\right]^{(1 + g^2_{\rm \scriptscriptstyle{MT}}
/\pi)^{-1}}}
\exp\left[\frac{2\pi}{\beta}X_1\left(1 + 
\frac{g^2_{\rm \scriptscriptstyle{MT}}}{\pi}\right)^{-1}\right] \, f_{+-}\left(z,\,\th,\, \sigma\,;X_1\right), \nn \\
\\
& & \la\la\,T_C\sigma_+(X)\sigma_+(0) \,\ra\ra \nn \\
&=& \left(\frac{T}{m_{\rm \scriptscriptstyle{MT}}}\right)^2
\left[4Q^2(X_1,X_2)\right]^{(1 + g^2_{\rm \scriptscriptstyle{MT}}
/\pi)^{-1}}
\exp\left[-\frac{2\pi}{\beta}X_1\left(1 + 
\frac{g^2_{\rm \scriptscriptstyle{MT}}}{\pi}\right)^{-1}\right]\,f_{++}(z,\,
\th,\,\sigma\,;X_1), \nn \\ 
\end{eqnarray}
where $z,\th$ and $\sigma$ are defined by 
\begin{eqnarray}
z=\frac{m^2_{\rm \scriptscriptstyle{MT}}}{2T}\left(\frac{2T}{m_{\rm \scriptscriptstyle{MT}}}\right)^{(1 + g^2_{\rm \scriptscriptstyle{MT}}/\pi)^{-1}},\quad 
\th = T,\quad \sigma = T\sqrt{\frac{\pi}{\pi + g^2_{\rm \scriptscriptstyle{MT}}}}.
\end{eqnarray}
Here $X=(X_1,X_2)$ is the coordinate on the two-dimensional Euclidean 
space, and $T_C$ denotes the contour-ordering along 
$C=[0,-i\beta]$ and
$Q^2(X_1,X_2)$ is defined by 
\begin{equation}
 Q^2(X_1,X_2)\;=\; \mathrm{sinh}\left(\frac{\pi}{\beta}(X_1 + iX_2)\right)
\mathrm{sinh}\left(\frac{\pi}{\beta}(X_1 - iX_2)\right).
\end{equation} 
Using these functions, we can define the scale-independent ratio 
\begin{equation}
 R(g_{\rm \scriptscriptstyle{MT}},T;X) \;=\; 
\frac{1}{\left[4Q^2(X_1,X_2)\right]^{2(1 + g^2_{\rm \scriptscriptstyle{MT}}
/\pi)^{-1}}}
\exp\left[\frac{4\pi}{\beta}X_1\left(1 + 
\frac{g^2_{\rm \scriptscriptstyle{MT}}}{\pi}\right)^{-1}\right]\,\frac{f_{+-}(z,\th,\sigma,;X)}{f_{--}(z,\th,\sigma ;X)} 
\end{equation}

Using the above quantities, we can also evaluate the condensate of a pair of 
the monopole and anti-monopole from the two-point function of the chiral
operators $\sigma_+$ and $\sigma_-$ by identifying
$\la\la\,\sigma_{+}\sigma_{-}\,\ra\ra\,(\,\la\la\,\sigma_{+}\sigma_{+}\,\ra\ra\,)$
with $\la\la\, M\bar{M}\,\ra\ra\,(\,\la\la\,MM\,\ra\ra\,)$. From the
results of the previous subsection we can read off the expression 
of $\la\la\, M\bar{M}\,\ra\ra\,(\,\la\la\,MM\,\ra\ra\,)$ plugged with
Eq.\,(\ref{MT-gauge}) as 
\begin{eqnarray}
& & \la\la\, M \bar{M} \,\ra\ra  \quad ( \;=\; \la\la\,  \bar{M} M \,\ra\ra    \;)\nn \\
&=& \frac{T^2}{2\zeta\Lambda^2}
 \frac{1}{\left[4Q^2(X_1,X_2)\right]^{2\pi^2/g^2}}
\exp\left[\frac{4\pi^3}{g^2}\frac{X_1}{\beta}\right] \, f_{+-}\left(z,\,\th,\, \sigma\,;X_1\right), \nn \\
\\
& & \la\la\,MM \,\ra\ra  \quad ( \;=\; \la\la\,  \bar{M} \bar{M} \,\ra\ra \;)   \nn \\
&=& \frac{T^2}{2\zeta^2\Lambda^2}
\left[4Q^2(X_1,X_2)\right]^{2\pi^2/g^2} 
\exp \left[ -\frac{4\pi^3}{g^2}\frac{X_1}{\beta} \right]\,
f_{++}(z,\,\th,\,\sigma\,;X_1), \nn \\ 
\end{eqnarray}
where $z,\th$ and $\sigma$ are defined by 
\begin{eqnarray}
z=\frac{\zeta\Lambda^2}{T}\left(\sqrt{\frac{2}{\zeta}}\frac{T}{\Lambda}
\right),\quad 
\th = T,\quad \sigma = \frac{\sqrt{2}\pi}{g}T.
\end{eqnarray}
The scale-independent ratio is given by
\begin{equation}
 R(g,\zeta,T;X) \;=\; 
\frac{1}{\left[4Q^2(X_1,X_2)\right]^{4\pi^2/g^2}}
\exp\left[\frac{8\pi^3}{g^2}\frac{X_1}{\beta}\right]\,
\frac{f_{+-}(z,\th,\sigma,;X)}{f_{--}(z,\th,\sigma ;X)}.  
\end{equation}

The neutral
pair condensate 
$\la\la\,M\bar{M}\,\ra\ra$ is much larger than $\la\la\,MM\,\ra\ra$ 
one at short distance (but yet
large distance compared to ($2\pi T)^{-1}$) as discussed in detail in
Ref.\,\cite{steer}. That is, the system tends to
form the $M\bar{M}$ pairs (``monopole bound states'') at high
temperature. Of course, the
behavior of those agrees with the physical expectation as noted in
Introduction. For large distance, no bound states are formed ($R$
tends to 1). The $R$ is almost 1 at the certain value $X_1$.

\section{Conclusions and Discussions}

We have investigated a phase structure of 
compact QED by considering the system as a perturbative deformation from
the topological model. Phases of compact QED are determined by the
behavior of the CG system in an $O(2)$NLSM$_2$. In this paper, in order
to discuss the behavior of the CG system, 
we have utilized a one-dimensional CG approximation and 
the thermal pressure that can be exactly calculated \cite{1CG}.
The thermal pressure of the one-dimensional CG system 
can be also expressed by parameters of compact QED.  
From the result we could study a phase structure of
compact QED with strong coupling and at high temperature. 
The critical-line equation has been explicitly evaluated. 
In particular, the asymptotic form of the critical-line 
equation derived in this paper is precisely identical with 
the result given by a \emph{perturbative} method (one-loop potential
calculation) in our previous work \cite{KY3}. A one-dimensional CG
approximation does not on a perturbation theory in two-dimensions, 
and hence this result is non-trivial and further confirmation
for the expression of the critical-line equations in the sufficiently
high-temperature and strong-coupling regime of compact QED. 
We have also discussed the relation between the chiral symmetry
restoration in the two-dimensional MT model and monopole condensate in
the four-dimensional compact QED by using some exact results in the
one-dimensional CG system. In particular, we have evaluated  
the monopole condensate in our scenario and discussed the quantum theory 
of monopoles from the viewpoint of the MT model. We might be able to 
relate the SG/MT duality in two dimensions to the electric/magnetic 
duality in four dimensions through the PS dimensional reduction. 
  
The two-dimensional MT model, SG model and CG system 
are not only interesting as the model in the statistical mechanics 
but also very useful to develop the techniques which may be applicable
to more realistic four-dimensional QCD as also 
noted in Ref.\,\cite{steer}. In particular, the MT model 
may be considered to be the toy model of QCD.
It should be remarked that at low temperature 
where the quarks and gluons are strongly confined into hadrons, one can
describe the system by an effective chiral bosonic Lagrangian (CBL) for
the lightest mesons (pions, kaons and eta) which are the Nambu Goldstone
bosons of the chiral symmetry breaking.
The relationship between the original fermionic QCD and the
chiral bosonic theory has many similarities with 
the duality between the SG and MT model. 

Moreover, it would possibly be related to the above analogy, 
compact QED is also interesting in the context of QCD with the
maximal Abelian gauge fixing. Once this gauge fixing is chosen, the full
QCD is thought to be described by compact $U(1)$ gauge theory in the
low energy (infrared) regime i.e., strongly-coupled regime. 
This phenomenon is called ``Abelian
dominance,'' confirmed by numerical simulations in the
lattice gauge theory. Thus compact QED is one of the most 
interesting subjects in the particle physics and 
well-suited to develop the techniques for studying the confinement in 
non-abelian gauge theories, such as QCD. 

Finally, we comment on the future problems. In our previous works
\cite{KY2,KY3}, we have obtained some plausible and consistent results 
but our considerations are restricted within the high-temperature and 
strong-coupling regime. For a example, the low-temperature regime 
near critical coupling is too difficult to study. However,
we expect that the low-temperature regime near the
critical coupling can be approached by the use of the Gaussian 
effective potential (GEP) that 
can describe a confining phase transition at zero temperature. 
We could study more intensively in this approach the critical behavior of 
compact QED at low temperature \cite{KY6}. 
Furthermore, we think that 
the recent results of Ref.\,\cite{FI} would be useful
to study the weak-coupling region of compact QED, $g < \pi$. 
This would be very interesting approach.

\vspace*{1cm}
\leftline{\bf\large Acknowledgements}
\vspace*{0.5cm}
\noindent
The author thanks 
W.~Souma, M.~Ishibashi and M.~Hamanaka for their encouragement, 
useful discussions and valuable comments.  
He also would like to acknowledge H.~Aoyama and K.~Sugiyama. 

\vspace*{1cm}

\appendix

\section{Critical Coupling from BKT Phase Transition}
\label{BKT:app}

Let us consider a pair of vortices at a finite distance $r_{12}$ in 
a box with
size $L$. From Eq.\,(\ref{CG-vol}) the global contribution of a vortex
pair to the partition function is, for dimensional reasons, of the form 
\begin{eqnarray}
 Z_{\rm pair} &\sim& \int_{r_{12}> R_0}\hspace{-0.8cm}d^2 x_1 d^2 y_1\,\exp\left(-\frac{4\pi^2}{g^2}\, \ln\frac{r_{12}}{R_0}\right) \nn \\
&\sim& L^4\exp\left(- \frac{4\pi^2}{g^2}\,\ln\frac{L}{R_0}\right).
\end{eqnarray}
Therefore we obtain the contribution to the free energy, 
\begin{equation}
 F_{\rm pair} \;\sim\; \ln Z_{\rm pair} \;\sim\; \left(4 - \frac{4\pi^2}{g^2}\right)\,\ln L.
\end{equation} 
If $g < \pi$, this is negligible in the limit $L\rar \infty$. While, if
$g > \pi$, the system becomes unstable. The well-separated vortices tend
to be created and the disorder increases. Thus we can estimate 
the critical-coupling as $g_{\rm cr} = \pi$. It corresponds to 
the critical temperature of the CG system $T_{\rm CG}= 1/8\pi$, 
at which the BKT phase transition occurs. 

\section{Finite-Size Effect of Cylinder}\label{pressure:app}

A massless free boson field theory 
is a conformal field theory with the central charge
$c=1$. Using the mapping from the infinite plane (with holomorphic
coordinate $z$) to the cylinder of circumference $\beta \equiv 1/T$ (with coordinate $w$), 
\begin{equation}
 w \;=\; \frac{\beta}{2\pi}\,\ln z~~(z=\e^{2\pi w/\beta}),
\end{equation}
we can obtain the free boson theory on the cylinder, and 
the free energy per unit length is shifted by 
\begin{equation}
 \Delta f \;=\; - \frac{\pi}{6}T^2,
\end{equation} 
due to the finite-size effect of the cylinder.  
Thus the free energy is shifted by 
\begin{equation}
 F \;=\; -\frac{\pi}{6}T^2 L, 
\end{equation}
and the partition function is given by 
\begin{equation}
 Z \;=\; \e^{-\beta F} \;=\; \exp\left(\frac{\pi}{6}LT\right).
\end{equation}
As a result, the additional pressure 
\begin{equation}
 P(T)\; =\; \frac{\pi}{6}T^2,
\end{equation}
arises from the geometry of the cylinder.

\section{One-Loop Effective Potential of the Sine-Gordon Model}

\begin{figure}
\begin{center}
 \includegraphics{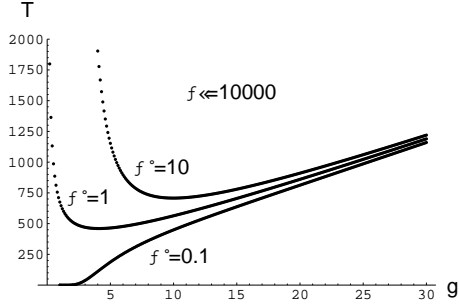}
\end{center}
\caption{\footnotesize The phase structure of a compact $U(1)$ gauge
 theory estimated by the one-loop effective potential of the SG model. } 
\label{plot:fig}
\end{figure}

The one-loop effective potential of the two-dimensional SG model at
finite temperature \cite{f-sine} 
is given by
\begin{eqnarray}
\label{4.1}
 V_{1{\rm loop}}(\phi_c) &=& V_{\rm 0}(\phi_c) + V_{\rm FT}(\phi_c),  \\
\label{4.2}
V_{\rm 0}(\phi_c) &\equiv& \frac{m^2}{8\pi}\cos\left(\frac{\sqrt{\lambda}}{m}\phi_c
\right)\left[1 - \ln\cos \left(\frac{\sqrt{\lambda}}{m}\phi_c \right)\right], \\
\label{4.3}
V_{\rm FT}(\phi_c) &\equiv& \frac{1}{\pi\beta^2}\int^{\infty}_{0}dx \ln\left[1 -
 \exp\left(- \sqrt{x^2 + M^2(\phi_c)\beta^2}\right)\right], \\
M^2 (\phi_c) &\equiv& m^2 \cos \left( \frac{\sqrt{\lambda}}{m} \phi_c  \right).
\end{eqnarray}
The second equation (\ref{4.2}) is the temperature-independent part and
corresponds to the zero temperature result. The third equation (\ref{4.3}) 
is the temperature-dependent part and describes the thermal effect. It 
vanishes in the zero temperature limit, $\beta \rar \infty$ .  
The minimum of the potential $\phi_c = 0$ is still
stable under one-loop quantum fluctuations at zero temperature.
 Taking the second derivative of Eq.\,(\ref{4.1}) with respect to $\phi_c$
 at $\phi_c =0$, one can evaluate the critical-line equation
 \cite{critical} as 
\begin{eqnarray}
 m^2(\beta) &=& m^2 + \left.\frac{\pd^2 V_{\rm FT}}{\pd\phi_c^{~2}}\right|_{\phi_c =0} \nn \\
&=& m^2 - m^2\frac{\bar{\lambda}}{2\pi}f(\bar{\beta})  = 0,
\end{eqnarray}
where $\bar{\lambda} \equiv \lambda /m^2$, $\bar{\beta} \equiv \beta m$ and $f(\bar{\beta})$ is defined by 
\begin{equation}
 f(\bar{\beta}) \equiv \int^{\infty}_0\!\!\! dx\, \frac{1}{\sqrt{\bar{\beta}^2 + x^2} 
\left[\exp\left(\sqrt{\bar{\beta}^2 + x^2} \right) - 1\right]}.
\end{equation}
Thus the critical-line equation is given by 
\begin{equation}
\label{cri}
 1 - \frac{\bar{\lambda}}{2\pi}f(\bar{\beta}) = 0.
\end{equation}
Parameters of the SG model can be expressed by those of 
compact QED and hence we obtain
\begin{equation}
 \bar{\beta} 
\;=\; \frac{4\pi^{3/2}}{gT}\,\zeta^{1/2}\Lambda,\qquad \bar{\lambda} = \frac{8\pi^{3}}{g^2}, 
\end{equation}
where one should remember that $\zeta$ is defined by
\begin{equation}
\zeta \equiv \kappa^{2\pi^2/g^2},\qquad \kappa \equiv \frac{R_0}{a}.
\end{equation}
Therefore the critical-line equation can be rewritten as  
\begin{equation}
\label{cr-eq}
 1 - \frac{4\pi^{2}}{g^2}f\left(\frac{4\pi^{3/2}\zeta^{1/2}}{gT}\right) = 0.
\end{equation}
The numerical plots of Eq.\,(\ref{cr-eq}) are shown 
in Fig.\,\ref{plot:fig}. 
The above (below) of the plotted-lines corresponds 
to a Coulomb (confining) phase. It should be noted that the validity of 
this derivation is restricted within the strong-coupling regime of 
compact QED.  

In the weak-coupling and high-temperature regime of the SG model, 
the critical-line Eq.\,(\ref{cri}) is reduced to the simple form,
\begin{equation}
\label{temp}
 T_{\rm cr} \;=\; \frac{4m^3}{\lambda},
\end{equation}
which denotes the critical-temperature of the SG model. 
Combining  Eq.\,(\ref{relation}) with Eq.\,(\ref{temp}) leads to 
\begin{equation}
\label{grad}
 T_{\rm cr} \;=\; \frac{2 \zeta^{1/2}\Lambda}{\pi^{3/2}}\,g \;=\; 
\frac{2 \Lambda \kappa^{\pi^2/g^2}}{\pi^{3/2}} \,g,
\end{equation}
which is the critical-temperature of compact QED with strong coupling 
and at high temperature.   

The SG model with weak coupling
corresponds to compact QED with strong one, and hence we can 
reliably evaluate the strong-coupling regime of compact QED 
since one-loop effective potential calculations are valid 
in the weakly coupled SG model. This fact is one of the greatest 
advantages in our scenario.

\end{document}